\documentclass[aps,prfluids,showpacs,notitlepage,superscriptaddress,longbibliography]{revtex4-2}
\usepackage{epsfig,graphics,amssymb,amsmath,subeqnarray,color,bm,bbm}
\usepackage[toc,page]{appendix}
\usepackage{mathtools}
\usepackage{xcolor}
\usepackage{float}
\usepackage{graphicx}
\usepackage[colorlinks=true,allcolors=blue]{hyperref}

\usepackage{mathrsfs}

\begin{document}
\title{A comprehensive Darcy-type law for viscoplastic fluids: I. Framework}
\author{Emad Chaparian}
\email{emad.chaparian@strath.ac.uk}
\affiliation{James Weir Fluid Laboratory, Department of Mechanical \& Aerospace Engineering, University of Strathclyde, Glasgow, United Kingdom}
\date{\today}

%%%%%%%%%%%%%%%%%%%%%%%%%%%%%%%%%%%%
%%%%%%%%%%%%%%%%%%%%%%%%%%%%%%%%%%%%
%%%%%%%%%%%%%%%%%%%%%%%%%%%%%%%%%%%%
%%%%%%%%%%%%%%%%%%%%%%%%%%%%%%%%%%%%
\begin{abstract}
A comprehensive Darcy-type law for viscoplastic fluids is proposed. Different regimes of yield-stress fluid flow in porous media can be categorised based on the Bingham number (i.e.~the ratio of the yield stress to the characteristic viscous stress). In a recent study (Chaparian, J.~Fluid Mech., vol.~980, A14, 2024), we addressed the yield/plastic limit (infinitely large Bingham number), namely, the onset of flow when the applied pressure gradient is just sufficient to overcome the yield stress resistance. A purely geometrical universal scale was derived for the non-dimensional critical pressure gradient, which was thoroughly validated against computational data. In the present work, we investigate the Newtonian limit (infinitely large pressure difference compared to the yield stress of the fluid --- ultra low Bingham number) both theoretically and computationally. We then propose a Darcy-type law applicable across the entire range of Bingham numbers by combining the mathematical models of the yield/plastic and Newtonian limits. Exhaustive computational data generated in this study (using augmented Lagrangian method coupled with anisotropic adaptive mesh at the pore scale) confirm the validity of the theoretical proposed law.
\end{abstract}
\maketitle
%%%%%%%%%%%%%%%%%%%%%%%%%%%%%%%%%%%%
%%%%%%%%%%%%%%%%%%%%%%%%%%%%%%%%%%%%
%%%%%%%%%%%%%%%%%%%%%%%%%%%%%%%%%%%%
%%%%%%%%%%%%%%%%%%%%%%%%%%%%%%%%%%%%
\section{Introduction}
%%%% Insert A head here

Fluid flow through porous media is a classical industrially relevant problem that has motivated extensive research over the years. Its importance spans a wide range of disciplines, including emerging applications such as CO$_2$ storage. Historically, the primary focus was on viscous fluid flows, particularly in contexts such as water resource management and hydrogeology. These early investigations led to proposing/deriving of the well-known Darcy's law \cite{darcy1856fontaines,whitaker1986flow}:

\begin{equation}\label{eq:Darcy_Newtonian}
\frac{\Delta \hat{P}}{\hat{L}} = \frac{\hat{\mu}}{\hat{\kappa}_0} \hat{U}_D,
\end{equation}
in which $\Delta \hat{P}/\hat{L}$ denotes the absolute value of the applied pressure gradient deriving the flow through the medium, $\hat{\mu}$ is the fluid viscosity, $\hat{\kappa}_0$ is a proportionality factor known as the {\it permeability}, and $\hat{U}_{D}$ represents the volume-averaged ``superficial'' velocity:
\begin{equation}\label{eq:superficialvelocity}
\hat{U}_{D} = \frac{\displaystyle \int_{\Omega \setminus \bar{X}} \hat{u} ~\text{d}\hat{V}}{ \hat{V}} = \frac{\displaystyle \hat{L} \int_{\Omega \setminus \bar{X}} \hat{u} ~\text{d}\hat{A}}{ \hat{A} \hat{L} } = \frac{\displaystyle \int_{\Omega \setminus \bar{X}} \hat{u} ~\text{d}\hat{A}}{ \hat{A}_{inl} } \frac{\hat{A}_{inl}}{ \hat{A} } = \hat{U} \frac{\hat{A}_{inl}}{ \hat{A} } = \frac{\hat{Q}}{\hat{A}},
\end{equation}
where $\hat{A}$ is the total cross-sectional area of the medium (i.e.~the area perpendicular to the flow direction), $\hat{A}_{inl}$ is the pore area at the inlet (i.e.~the area through which the fluid flows or can potentially flow), $\hat{V}$ is the total volume of the domain $\Omega$ (i.e.~$\hat{V}=\text{meas}(\Omega)$) which includes both the solid obstacles/voids (i.e.~$X$) and the pore space (i.e.~$\Omega \setminus \bar{X}$), $\hat{U}$ represents the mean fluid velocity at the inlet and $\hat{Q}$ is the volumetric flow rate. In some references, the superficial velocity is alternatively defined as $\hat{U}_D = (1-\phi) ~\hat{U}$ for stochastically homogeneous porous media. Here, $\phi$ denotes the obstructed volume fraction (i.e.~$\phi=\text{meas}(X)/\text{meas}(\Omega)$) which corresponds to the portion of the domain where fluid cannot flow, while $1-\phi$ is the porosity of the media. For viscous fluids, Darcy’s law (\ref{eq:Darcy_Newtonian}) predicts a linear relationship between the volumetric flow rate and the applied pressure gradient, with the permeability depending solely on the geometry and structure of the medium.

Beyond such applications, there are numerous cases in which non-Newtonian fluids flow through porous media. A very important class of practical fluids is the yield-stress fluids which are intrinsic to many industrial and biomedical applications. Examples include enhanced oil recovery \cite{wang2023review}, fracture/formatrion stabilisation and cement grouting \cite{Ulf2021,da2023comprehensive}, squeeze cementing \cite{izadi2023}, vertebroplasty and cementoplasty \cite{trivedi2023continuum}, etc. These materials are neither ideal solids nor ideal liquids: they behave like solids when the applied stress is less than a threshold --- referred to as the yield stress ($\hat{\tau}_y$) --- and flow like fluids when the applied stress exceeds this threshold. A wide range of rheological models has been proposed to describe the behaviour of such materials. The simplest class is the viscoplastic models: the Bingham model which assumes a constant plastic viscosity ($\hat{\mu} = \hat{\mu}_p$) and the Herschel–Bulkley model which incorporates shear-dependent viscosity governed by a consistency parameter $\hat{K}$ and a power-law index $n$. The viscosity functional in the latter is given by $\hat{\mu} = \hat{K} \hat{\dot{\gamma}}^{n-1}$, where $\hat{\dot{\gamma}}$ is the shear rate.

\begin{figure}
\begin{center}
\includegraphics[width=0.92\textwidth]{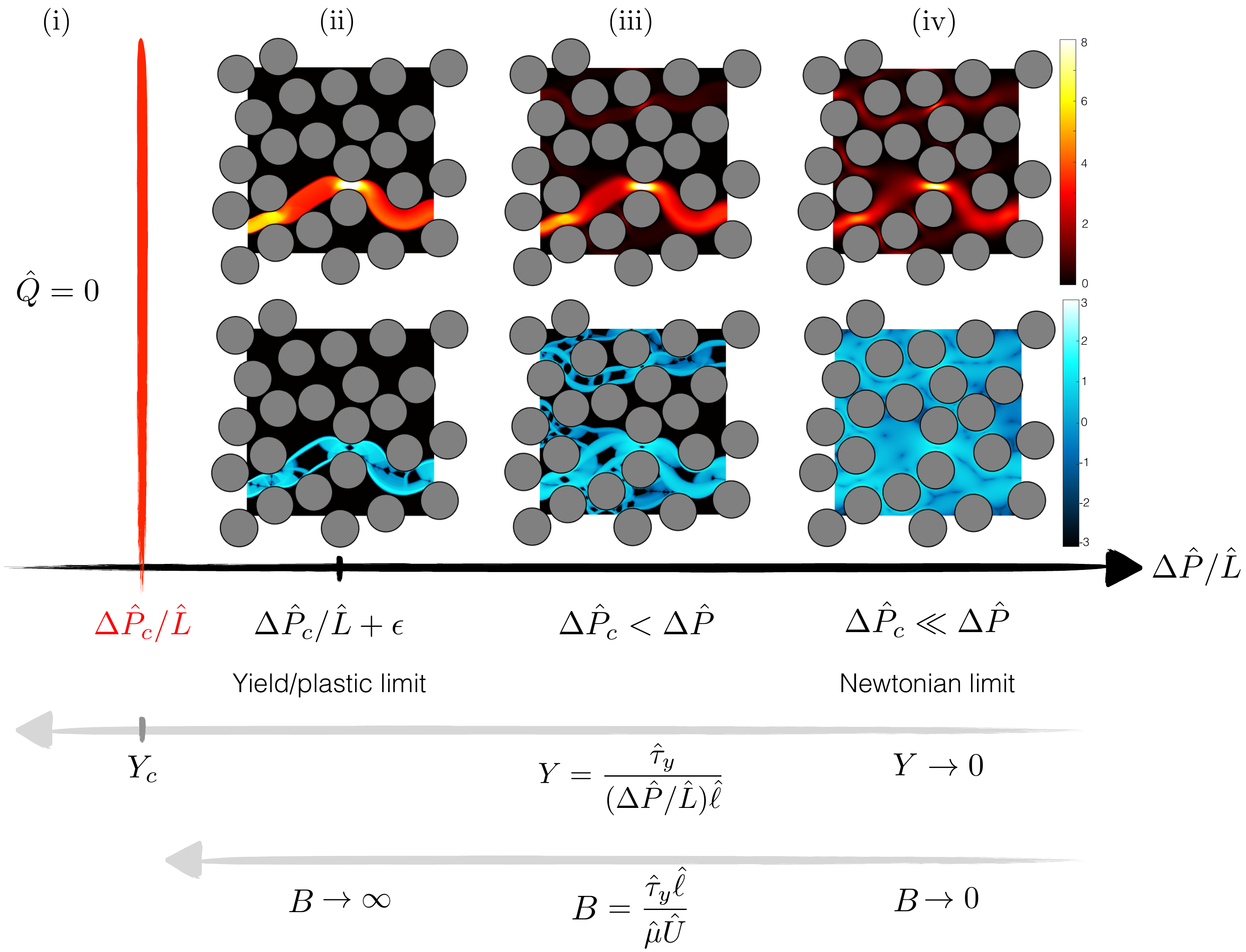}
\caption{Categorisation of yield-stress fluid flows in a porous medium (made of cylindrical obstacles - $\phi \approx 0.5$): (i) when the bulk pressure gradient is less than the critical value $\Delta \hat{P}_c/\hat{L}$, the flow rate is zero and the fluid remains entirely unyielded; (ii) just above the critical threshold, flow initiates but remains highly localised within a single channel; (iii) as the bulk pressure gradient increases, more channels/pathways appears and the flow rate increases accordingly; (iv) when the bulk pressure gradient is much larger than the critical value, the fluid is nearly yielded everywhere. The top row shows the contours of the non-dimensional velocity, while the bottom row displays $\log_{10}(\Vert \dot{\boldsymbol{\gamma}} \Vert)$ for a Bingham fluid at $B=0,~10^2$ and $10^4$, from right to left. Note that all panels within each row share the same colour bar range. The additional axes illustrate the non-dimensional yield and Bingham numbers.}
\label{fig:regimes}
\end{center}
\end{figure}

In the case of yield-stress fluids flowing through porous media, different flow regimes can be identified based on the balance between the yield stress and the applied pressure difference (or equivalently the characteristic viscous stress) see Fig.~\ref{fig:regimes}. Due to the presence of a yield stress, there exists a critical pressure gradient ($\Delta \hat{P}_c/\hat{L}$) that must be exceeded to start/sustain the flow. When the applied pressure gradient is just sufficient to initiate the flow, the fluid motion is highly localised in a single path \cite{talon2013determination,liu2019darcy,chaparian2021sliding,fraggedakis2021first,chaparian2024yielding} --- plastic/yield limit. As the applied pressure gradient increases, a larger portion of the fluid yields and flows through the porous structure. The ultimate regime is when the applied pressure gradient is much greater than the yield stress resistance resulting in the fluid being yielded and flowing `nearly' everywhere --- Newtonian/viscous limit.

Studying of how yield-stress fluids would flow in porous media dates back at least to 1960s, when extraction of heavy crude oil was a primary motivation. An interesting review of early works in this area can be found in \cite{frigaard2017bingham}. Since then, many studies have explored yield-stress fluid flows from various perspectives: pore scale characteristics (e.g.~\cite{waisbord2019anomalous}), role of complex rheologies such as elastoviscoplasticity (e.g.~\cite{chaparian2019porous,parvar2024general}) and intricate hydrodynamic phenomena such as fluid sliding over solid topologies (e.g.~\cite{chaparian2021sliding}). However, efforts to develop a macroscopic closure or a general Darcy-type law (which is the central focus of the present study) remain relatively limited. One of the earliest attempts in this direction was made by Al-Fariss \& Pinder \cite{alfariss1987flow}, who conducted a set of experiments in packed beds of sand with paraffin wax-in-oil solutions. Their measurements were also compared with a semi-empirical model derived by trying to generalise the Blake-Carman-Kozney expression for yield-stress fluids. An irrational assumption in the initial derivation steps (namely, the Poiseuille flow in a capillary tube) led to a model linearly composed of two terms:
\begin{equation}\label{eq:AlFariss_Pinder}
\frac{\Delta \hat{P}}{\hat{L}} = \frac{\hat{K}}{A(n)} \frac{(1-\phi)^{(1-n)/2}}{\hat{\kappa}_0^{(1+n)/2}} \hat{U}^n_D + \sqrt{\frac{25(1-\phi)}{24 \hat{\kappa}_0}} \hat{\tau}_y.
\end{equation}
The first term represents characteristic viscous stress, while the second term accounts for the yield stress. The key flaw in the model proposed by Al-Fariss \& Pinder lies in the assumption that, within a narrow conduit, the fluid is sheared/yielded throughout the entire cross section. This assumption is invalid, particularly in the case of non-uniform conduits, as has been previously demonstrated both theoretically/computationally \cite{hewitt2016heleshaw,izadi2023} and experimentally \cite{daneshi2020obstructed}. A more detailed discussion of this issue is provided in Appendices \ref{sec:appA} \& \ref{sec:appB}. In expression (\ref{eq:AlFariss_Pinder}), $A(n)$ is a function of the power-law index and reduces to unity for a Bingham fluid, i.e. when $n = 1$. As is clear, the use of this semi-empirical model requires prior knowledge of the permeability of the medium for a Newtonian (purely viscous) fluid, denoted by $\hat{\kappa}_0$. Nearly all succeeding works (focused on proposing a general Darcy-type law for yield-stress fluids) proposed expressions of the same type that can be condensed into the general form,
\begin{equation}\label{eq:generalDarcy}
\frac{\Delta \hat{P}}{\hat{L}} = \hat{\alpha} ~\hat{\mu} \left( \frac{\hat{U}_D}{\hat{\ell}} \right) + \hat{\beta} ~\hat{\tau}_y.
\end{equation}
In this expression, $\hat{\ell}$ is the characteristic length and $\hat{\alpha}$ \& $\hat{\beta}$ are unknown coefficients that must be determined by fitting to either numerical or experimental data. Examples are studies by Chevalier et al.~\cite{chevalier2013darcy,chevalier2014breaking} in which experimental measurements were performed on the flow of the water-in-oil emulsion through glass beads and by Bleyer \& Coussot \cite{bleyer2014breakage} where 2D numerical simulations of Herschel-Bulkley fluid through cylindrical obstacles with hexagonal arrangements were done. Shahsavari \& McKinley \cite{shahsavari2016mobility}, building on their earlier work on shear-thinning fluids \cite{shahsavari2015mobility}, took a deeper dive into the problem of 2D viscoplastic fluid flows through cylindrical obstacles with square arrangements. They proposed a rescaling based on the effective viscosity to propose a Darcy-type law which was similar to expression (\ref{eq:generalDarcy}). Another interesting series of studies on this matter have been done by Talon and collaborators \cite{talon2013determination,chevalier2015generalization,bauer2019experimental}, who push the theoretical analysis further by showing that $\hat{\alpha}$ \& $\hat{\beta}$ are independent of both the flow rate and the applied pressure gradient. However, these coefficients were not explicitly evaluated in their studies. In a recent work, Casta{\~n}eda \cite{castaneda2023variational} adopted a homogenisation approach (referred to as the ``variational linear comparison'') and derived an upper bound for the applied pressure gradient,
\begin{equation}\label{eq:castaneda}
\frac{\Delta \hat{P}}{\hat{L}} \leqslant \hat{K} \frac{(1-\phi)^{(1-n)/2}}{\hat{\kappa}_0^{(1+n)/2}} \hat{U}_D^n + \sqrt{\frac{1-\phi}{\hat{\kappa}_0}} \hat{\tau}_y.
\end{equation}
The form of this upper bound is very close to Al-Fariss \& Pinder's model (i.e.~expression (\ref{eq:AlFariss_Pinder})), yet its significance lies in the fact that it is derived from a rigorous theoretical framework, rather than a semi-empirical model. 

As is evident, all previously proposed models require either {\it a priori} knowledge of the medium's permeability in the Newtonian case ($\hat{\kappa}_0$) or unknown pre-factors. In the present study, we rather adopt a different approach to derive a general Darcy-type law for viscoplastic fluids. We begin by reviewing the Newtonian/viscous limit of the problem (i.e.~$\hat{\tau}_y \ll \hat{\mu}  ~\hat{U}_D / \hat{\ell} $ or $\Delta \hat{P}_c / \hat{L} \ll \Delta \hat{P} / \hat{L}$) in section \ref{sec:viscous}. We then combine this with the yield limit model (i.e.~$\hat{\tau}_y \gg \hat{\mu}  ~\hat{U}_D / \hat{\ell} $ or $\Delta \hat{P} / \hat{L} \to (\Delta \hat{P}_c / \hat{L})^+ $), which we previously developed in \cite{chaparian2024yielding} and revisit in section \ref{sec:yield}. The resulting general Darcy-type law is subsequently validated in section \ref{sec:darcy} by exhaustive numerical simulations. These are carried out using an adaptive finite element method at the pore scale based on an augmented Lagrangian approach \cite{roquet2003adaptive,chaparian2019adaptive,chaparian2022vane,chaparian2024yielding} over a wide range of parameters, including the intermediate regime (i.e.~$\hat{\tau}_y \sim \hat{\mu}  ~\hat{U}_D / \hat{\ell}$ ).

%%%%%%%%%%%%%%% End of first page %%%%%%%%%%%%%%%%%%%%%

\section{Revisiting viscous fluid flows in porous media}\label{sec:viscous}

As mentioned earlier, in this section, we review several models (among many others) proposed over the years to calculate the permeability of a medium to a viscous fluid, $\hat{\kappa}_0$. Our focus is more on models with a rather fundamental basis that can be directly applied for estimating $\hat{\kappa}_0$. While upper bounds for permeability have been derived (e.g.~by Prager \& Weissberg \cite{prager1,prager2,prager3} and Doi \cite{doi1976new}) using variational formulations with correlation functions and admissible stress fields, these approaches are not directly relevant to our current analysis and therefore are not included in this review. Critical reviews of these bounds can be found in \cite{berryman1985bounds} or most recently in \cite{bignonnet2018upper}.

\subsection{Blake-Carman-Kozeny (BCK): a semi-empirical model}\label{sec:BCK}
Perhaps the most well-known and widely used model for estimating permeability is the Blake-Carman-Kozeny (BCK) model. In the literature, it is also referred to as the Blake-Kozeny \cite{macdonald1991generalized} or the Kozney-Carman \cite{guyon2015physical} model. The derivation of this model is based on the simple Poiseuille flow,
\begin{equation}
\hat{Q}=\frac{\Delta \hat{P}}{\hat{L}} \frac{\pi \hat{R}_c^4}{8 \hat{\mu}} \rightarrow \hat{U} \sim \hat{R}^2_h \frac{\Delta \hat{P}}{\hat{L}},
\end{equation}
where $\hat{R}_c$ is the capillary radius and $\hat{R}_h$ denotes the well-known {\it hydraulic radius}. Now, considering the presence of spherical obstacles in the flow domain, and using Eqs.~(\ref{eq:Darcy_Newtonian}) \& (\ref{eq:superficialvelocity}) along with the above equation, we can deduce that,
\begin{equation}\label{eq:BCK_derivation}
\hat{U}_D = (1-\phi) \hat{U} = \frac{\hat{\kappa}_0}{\hat{\mu}} \frac{\Delta \hat{P}}{\hat{L}} \sim (1-\phi) \hat{R}^2_h \frac{\Delta \hat{P}}{\hat{L}},
\end{equation}
or indeed,
\begin{equation}
\hat{\kappa}_0 \sim (1-\phi) \hat{R}_h^2.
\end{equation}
By definition, $\hat{R}_h = \hat{V}_f / \hat{A}_{w}$ where $\hat{V}_f$ is the volume of the fluid (i.e.~$\hat{V}_f=\text{meas}(\Omega \setminus \bar{X})$) and $\hat{A}_w$ is the ``wetted area''. Considering a domain of total volume $\hat{V}$, the solid volume fraction can be expressed as $\phi=N \frac{(4/3) \pi \hat{R}^3}{\hat{V}}$ where $N$ is the number of non-overlapping spheres of radius $\hat{R}$. Consequently, the hydraulic radius can be written as,
\begin{equation}
\hat{R}_h = \frac{\hat{V} - \frac{4}{3} \pi \hat{R}^3 N}{4 \pi \hat{R}^2 N} \sim \frac{1-\phi}{\phi/\hat{R}}.
\end{equation}
Therefore, the permeability scale will be,
\begin{equation}\label{eq:BCK_scaling}
\frac{\hat{\kappa}_0}{\hat{R}^2} \sim \frac{(1-\phi)^3}{\phi^2}.
\end{equation}
Zick \& Homsy \cite{zick1982stokes} performed numerical simulations for simple, body-centred and face-centred cubic packing of spheres, demonstrating that the BCK scaling provides reasonably accurate predictions within the solid volume fraction range $\phi \in (0.5,0.75)$.

\subsection{Brinkman's model}
Brinkman \cite{brinkman1947} further examined the BCK model and identified a key limitation: in the dilute regime ($\phi \to 0$), where the system consists of widely spaced individual spherical obstacles, the BCK model predicts a diverging drag on a single sphere. This contradicts Stokes' drag law for a sphere. Suggesting a correction to the {\it local} Darcy law,
\begin{equation}
\boldsymbol{\nabla} \hat{p} = \frac{\hat{\mu}}{\hat{\kappa}_0} \hat{\boldsymbol{u}} + \hat{\mu}^{\prime} \boldsymbol{\nabla}^2 \hat{\boldsymbol{u}},
\end{equation}
Brinkman defined an ``artificial viscosity'', $\hat{\mu}^{\prime}$, to account for the hydrodynamic interactions of the fields around each sphere. In the above equation, $\hat{\boldsymbol{u}}$ is the local fluid velocity vector. After some algebra, Brinkman derived,
\begin{equation}
\frac{6\pi \hat{\mu} \hat{R} \hat{U}_{\infty}N}{(\Delta \hat{P}/\hat{L}) \hat{V}} = \frac{\hat{\mu}^{\prime}}{\hat{\mu}} \left[ 1+ \frac{3 \phi}{4} \left( 1- \sqrt{\frac{8}{\phi}-3} \right) \right]
\end{equation}
where $\hat{U}_{\infty}$ denotes the fluid velocity far from the spheres. This expression asymptotically reduces to the Stokes' drag law in the dilute limit $\phi \to 0$. For the viscosity ratio $\hat{\mu}^{\prime} / \hat{\mu}$, Brinkman suggested adopting Einstein's formula for dilute suspensions (i.e.~$1 + 2.5 \phi$) or determining it empirically by fitting to experimental data. In the leading order, and as originally proposed by Brinkman, the viscosity ratio can be approximated by unity, which leads to,
\begin{equation}
\frac{\hat{\kappa}_0}{\hat{R}^2} \sim \frac{1}{\phi} + \frac{3}{4} \left( 1 - \sqrt{\frac{8}{\phi} -3} \right).
\end{equation}
Cancelliere et al. \cite{cancelliere1990permeability} performed numerical simulations of fluid flow through porous media composed of randomly shaped obstacles and concluded that the Brinkman model provides better accuracy at low solid volume fractions (i.e., $\phi \in [0, 0.1)$), whereas the BCK model performs more reliably at higher volume fractions (i.e., $\phi \in [0.25, 1)$); see figure 2 of the reference.

\subsection{Series-based model}\label{sec:series}
Another class of theoretical models for Newtonian permeability is based on periodic series solutions to the Stokes equation. For instance, Hasimoto \cite{hasimoto1959periodic} analysed flow over an array of spheres by expanding the velocity and pressure gradient by the Fourier series over a unit lattice of spheres/obstacles. To satisfy the boundary conditions on individual obstacles, Hasimoto employed an analogy with the problem of ionic crystals and electrostatic potentials which is valid for small spheres only. Then, he modified the original approximation for various arrangements such as simple, body-centred and face-centred cubic packings, with all converging to the Stokes' drag law on a sphere in the limit of $\phi \to 0$. A comprehensive historical review of such models, particularly in 2D based on Oseen’s equations and the drag on a cylinder, is provided in \cite{drummond1984laminar}. These models typically share the compact form:
\begin{equation}\label{eq:series_drag}
\hat{F} = \frac{4\pi \hat{\mu} \hat{U}}{-\log(\hat{R} / \hat{L}) + b + c (\hat{R} / \hat{L}) + O((\hat{R} / \hat{L})^2)},
\end{equation}
where $\hat{F}$ is the drag force per unit length on the cylinders, which can be converted to permeability and/or pressure drop across the porous medium. We will employ this class of models in section \ref{sec:darcy} to propose a general Darcy-type law for viscoplastic fluids.

\section{Modelling yield-stress fluid flows through porous media}\label{sec:math}

\subsection{Mathematical formulation}

We consider incompressible two-dimensional Stokes flow in porous media (i.e.~flow over a set of obstacles) which is governed by,
\begin{equation}\label{eq:Stokes_dimensional}
\boldsymbol{0} = - \boldsymbol{\nabla} \hat{p} + \boldsymbol{\nabla} \boldsymbol{\cdot} \boldsymbol{\hat{\tau}} ~~\&~~\boldsymbol{\nabla} \boldsymbol{\cdot} \hat{\boldsymbol{u}} = 0,~~~\text{in}~ \Omega \setminus \bar{X},
\end{equation}
where $\hat{\boldsymbol{u}} = (\hat{u},\hat{v})$ is the velocity vector and $\boldsymbol{\hat{\tau}}$ is the deviatoric stress tensor. The Cauchy stress tensor can be defined as $\boldsymbol{\hat{\sigma}}=-\hat{p} \boldsymbol{1} + \boldsymbol{\hat{\tau}} $ where $\boldsymbol{1}$ is the identity matrix.

To describe the rheological behaviour of a viscoplastic fluid, we use the Bignham model in the present study,
\begin{equation}\label{eq:const_dimensional}
  \left\{
    \begin{array}{ll}
      \hat{\boldsymbol{\tau}} = \left( \hat{\mu}_p + \displaystyle{\frac{\hat{\tau}_y}{\Vert \hat{\dot{\boldsymbol{\gamma}}} \Vert}} \right) \hat{\dot{\boldsymbol{\gamma}}} & \mbox{iff}\quad \Vert \hat{\boldsymbol{\tau}} \Vert > \hat{\tau}_y, \\[2pt]
      \hat{\dot{\boldsymbol{\gamma}}} = 0 & \mbox{iff}\quad \Vert \hat{\boldsymbol{\tau}} \Vert \leqslant \hat{\tau}_y,
  \end{array} \right.
\end{equation}
where $\hat{\dot{\boldsymbol{\gamma}}}$ is the rate-of-strain tensor or indeed $\boldsymbol{\nabla} \hat{\boldsymbol{u}} + \boldsymbol{\nabla} \hat{\boldsymbol{u}}^T$ and $\Vert \cdot \Vert$ is the second invariant of a tensor (i.e.~$\Vert \boldsymbol{\Lambda} \Vert = \sqrt{(1/2) \boldsymbol{\Lambda} \boldsymbol{:} \boldsymbol{\Lambda}}$ when $\boldsymbol{\Lambda}$ is deviatoric). Therefore, the von Mises yielding criterion is considered for the fluid.

To scale the Eqs.~(\ref{eq:Stokes_dimensional}) \& (\ref{eq:const_dimensional}), we follow what is known as the ``resistance formulation'' (see \cite{chaparian2021sliding,chaparian2024yielding} for details). Indeed, the velocity vector is scaled with the mean velocity at the inlet, $\hat{U}$, and the deviatoric stress tensor/pressure with the characteristic viscous stress, $\hat{\mu}_p \hat{U}/\hat{\ell}$ where $\hat{\ell}$ is the length scale which will be fixed later in this section. Hence, the non-dimensional forms of the governing and constitutive equations read, 
\begin{equation}\label{eq:Stokes}
0 = - \boldsymbol{\nabla} p + \boldsymbol{\nabla} \boldsymbol{\cdot} \boldsymbol{\tau} ~~\&~~\boldsymbol{\nabla} \boldsymbol{\cdot} \boldsymbol{u} = 0,~~~\text{in}~ \Omega \setminus \bar{X},
\end{equation}
and,
\begin{equation}\label{eq:const}
  \left\{
    \begin{array}{ll}
      \boldsymbol{\tau} = \left( 1 + \displaystyle{\frac{B}{\Vert \dot{\boldsymbol{\gamma}} \Vert}} \right) \dot{\boldsymbol{\gamma}} & \mbox{iff}\quad \Vert \boldsymbol{\tau} \Vert > B, \\[2pt]
      \dot{\boldsymbol{\gamma}} = 0 & \mbox{iff}\quad \Vert \boldsymbol{\tau} \Vert \leqslant B,
  \end{array} \right.
\end{equation}
where $B = \hat{\tau}_y \hat{\ell} / \hat{\mu}_p \hat{U} $ is the Bingham number, representing the ratio of the yield stress to the characteristic viscous stress. For dimensionless quantities, the {\it hat} sign ($\hat{\cdot}$) is dropped. In this type of scaling, for a fixed geometry, the only independent parameter is the Bingham number, since the non-dimensional flow rate is fixed at $Q=L_{inl}$ (see expression (\ref{eq:superficialvelocity})). The bulk pressure drop can therefore be computed as a function of the Bingham number using the energy balance equation,
\begin{equation}\label{eq:energy}
a(\boldsymbol{u},\boldsymbol{u}) + B j(\boldsymbol{u}) = \int_{\Omega \setminus \bar{X}} (\dot{\boldsymbol{\gamma}} \boldsymbol{:} \dot{\boldsymbol{\gamma}} ) ~\text{d}A + B \int_{\Omega \setminus \bar{X}} \Vert \dot{\boldsymbol{\gamma}} \Vert ~\text{d}A = \frac{\Delta P}{L} \int_{\Omega \setminus \bar{X}} u ~\text{d}A,
\end{equation}
where $a(\boldsymbol{u},\boldsymbol{u})$ and $B j(\boldsymbol{u})$ represent the viscous and plastic dissipation functions, respectively. The permeability for a viscoplastic fluid flowing through a porous medium is then given by,
\begin{equation}\label{eq:}
\hat{\kappa} = \frac{\hat{\mu} ~\hat{U}_D}{\Delta \hat{P} / \hat{L}}  \Rightarrow \kappa = \frac{\hat{L}_{inl} / \hat{L}}{\Delta P/L}.
\end{equation}

As illustrated in Fig.~\ref{fig:regimes}, the Newtonian limit is recovered when $B \to 0$, leading to $\kappa \to \kappa_0$. In contrast, the yield limit is approached as $B \to \infty$, where $\kappa \to 0$.

\subsection{Porous media construction \& computational methodology}
To construct the porous media and solve the governing \& constitutive equations, we follow the same methodology fully detailed in our previous work \cite{chaparian2024yielding}. Just to highlight, the porous media construction in the present study is based on a randomised non-overlapping obstacle distribution (i.e.~$X$) in a $L \times L = 50 \times 50$ square domain $\Omega$. No constraints are imposed on the obstacle positions and they are allowed to cross the boundaries of the computational domain. In this study, we focus exclusively on square monodispersed obstacles with edge length $\sqrt{\pi}$. Indeed, the characteristic length is the radius of an equivalent circular obstacle, making the non-dimensional area of each square obstacle equal to $\pi$. As we are concerned with two-dimensional flows, the solid ``volume'' fraction is interpreted as the ratio of the total area occupied by the obstacles to the area of the domain $\Omega$.

Regarding numerical simulations, we implement augmented Lagrangian method to simulate viscoplastic fluid flows \cite{roquet2003adaptive}. This method effectively handles non-differentiable viscoplastic models by relaxing the rate-of-strain tensor. An open source finite element environment --- FreeFEM++ \cite{freefem} --- is used to discretise Eqs.~(\ref{eq:Stokes}) \& (\ref{eq:const}). Anisotropic adaptive meshing at the pore scale in $\Omega/\bar{X}$ is coupled with our computational scheme, see figure 2 of \cite{chaparian2024yielding}. Our entire computational framework has been extensively validated in our previous studies; for more details, see \cite{chaparian2017yield,chaparian2019adaptive,iglesias2020computing,chaparian2021sliding,chaparian2022vane,medina2023rheo}. The velocity boundary conditions on the faces of the computational domain and the enforcement of the flow rate are explained in \cite{chaparian2024yielding,chaparian2021sliding,chaparian2020stability}. A no-slip boundary condition is enforced on all solid obstacles.

\begin{figure}
\begin{center}
\includegraphics[width=0.75\textwidth]{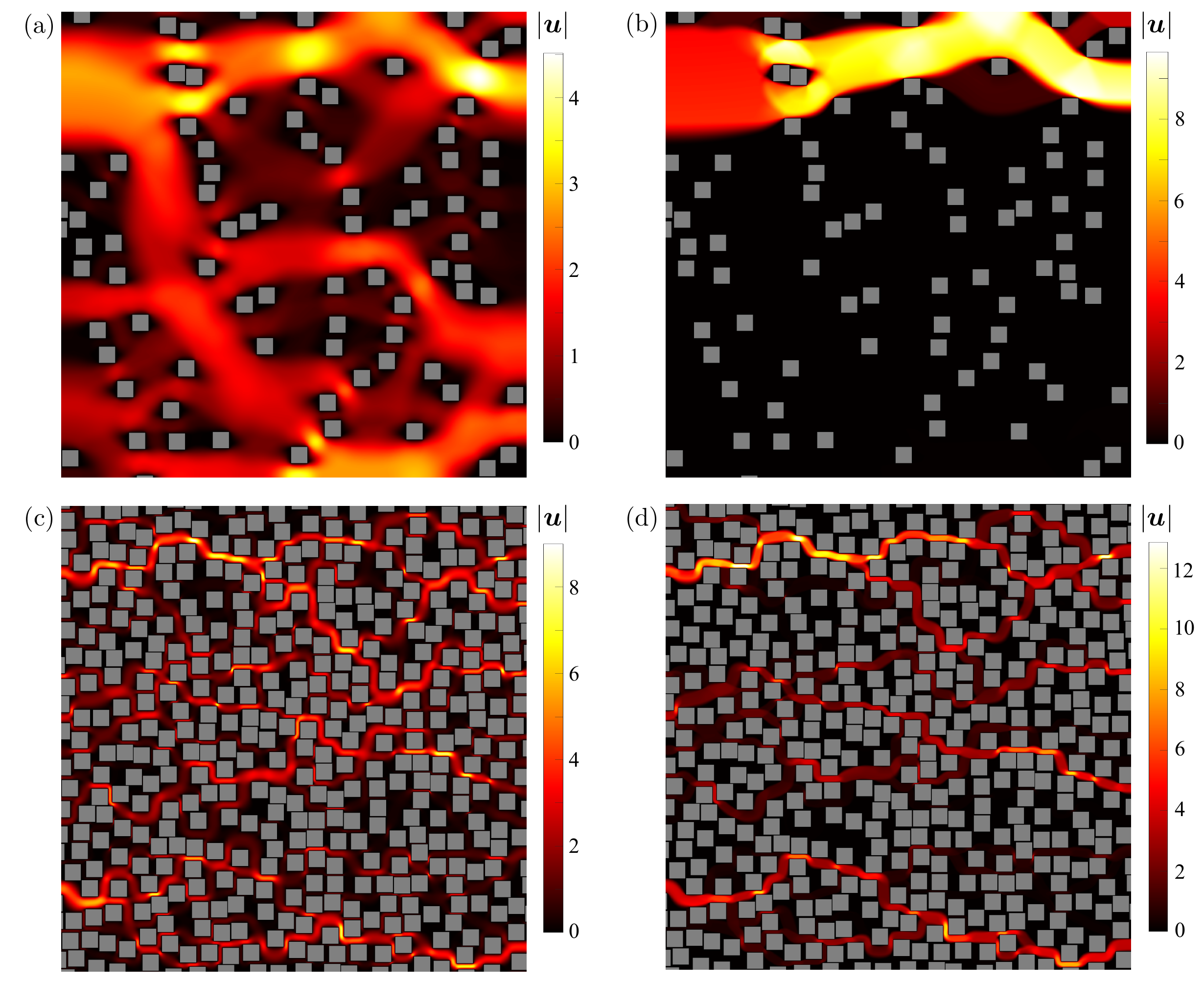}
\caption{Contours of velocity magnitude for two sample realisations: (a,c) $B=10^0$, (b,d) $B=10^2$. In top row, $\phi=0.1$ and in the bottom row, $\phi=0.5$.}
\label{fig:samples}
\end{center}
\end{figure}

Fig.~\ref{fig:samples} illustrates four sample simulations at $B=1$ (left column) and $B=100$ (right column) for two different media at $\phi=0.1$ (top row) and $\phi=0.5$ (bottom row).

\section{Yielding to flow}\label{sec:yield}
In this section, we examine the yield limit ($B \to \infty$) by revisiting our previously proposed model and the associated scaling derived in \cite{chaparian2024yielding}. Furthermore, we compare this scaling with another recently proposed upper bound in \cite{castaneda2023variational}. To facilitate the discussion, we define the yield number as,
\begin{equation}
Y = \frac{\hat{\tau}_y}{\left( \Delta \hat{P} / \hat{L} \right) \hat{\ell}} = \frac{B}{\Delta P/L}.
\end{equation}
The yield number is essentially the inverse of the non-dimensional applied pressure gradient. When the applied pressure gradient is less than the critical threshold ($\Delta \hat{P}_c / \hat{L}$), flow does not initiate, and $Y>Y_c$ indicates that the yield stress resistance is greater than the driving stress, i.e.~the system is completely unyielded with $\hat{Q}=0$. If $Y<Y_c$, however, flow occurs at least through the first open path, (see Fig.~\ref{fig:regimes}), leading to a non-zero flow rate. Hence, the critical yield number $Y_c$ characterises the onset of the flow. It can be calculated using our computational scheme via,
\begin{equation}\label{eq:Yc}
Y_c = \frac{\hat{\tau}_y}{\left( \Delta \hat{P}_c / \hat{L} \right) \hat{\ell}} = \lim_{B \to \infty} \frac{B}{\Delta P/L} = \lim_{B \to \infty} \frac{\displaystyle\int_{\Omega \setminus \bar{X}} u ~\text{d}A}{j(\boldsymbol{u})},
\end{equation}
which is indeed a rearrangement of expression (\ref{eq:energy}) considering that, at the yield limit, the viscous dissipation is at least one order of magnitude less than the plastic dissipation. In our prior work \cite{chaparian2024yielding}, we proposed scaling,
\begin{equation}\label{eq:h_L}
Y_c \sim \frac{ \bar{h}_{ch} }{ L_{ch} / L },
\end{equation}
for a porous medium, in which $h_{ch}$ is the local varying height of the first open path and $L_{ch} / L$ represents its {\it tortuosity}; see figure 5 of \cite{chaparian2024yielding}. In the above expression, $\bar{h}_{ch}$ represents the mean height over the total length of the first open channel. Using our extensive pore-network computational data of the randomised circular obstacles \cite{fraggedakis2021first}, we found that,
\begin{equation}\label{eq:scales}
\left< \bar{h}_{ch} \right> \sim 1-\phi ~~ \& ~~ \left< L_{ch} / L \right> \sim \phi,
\end{equation}
where $\left< \cdot \right>$ is the ensemble average over the entire set of simulations ($\approx 500$ simulations for each $\phi$). Substituting these into expression (\ref{eq:h_L}), we proposed,
\begin{equation}\label{eq:emad}
Y_c = \frac{1}{\pi} \frac{1-\phi}{\phi},
\end{equation}
which was validated across various types of porous media, including mono-dispersed and bi-dispersed circular, square, and polygon obstacles in \cite{chaparian2024yielding} (see figure 6 of the reference).

As mentioned, very recently, Casta{\~n}eda \cite{castaneda2023variational} derived an upper bound of the applied pressure gradient, expression (\ref{eq:castaneda}). In the yield limit, the leading order is the second term on the right hand side of (\ref{eq:castaneda}) as $\hat{U}_D \to 0$ in this limit, allowing us to convert it to an inequality,
\begin{equation}\label{eq:castaneda_yield}
\sqrt{\frac{\kappa_0}{(1-\phi)}} \leqslant Y_c.
\end{equation}
It is worth noting that using expression (\ref{eq:AlFariss_Pinder}) proposed by Al-Fariss \& Pinder \cite{alfariss1987flow} results in almost the same critical yield number (though with an equality sign), differing only by a factor of $\sqrt{24/25} \approx 0.98$. An alternative simple derivation presented in Appendix \ref{sec:appB} yields the same scaling.

\begin{figure}
\begin{center}
\includegraphics[width=0.8\textwidth]{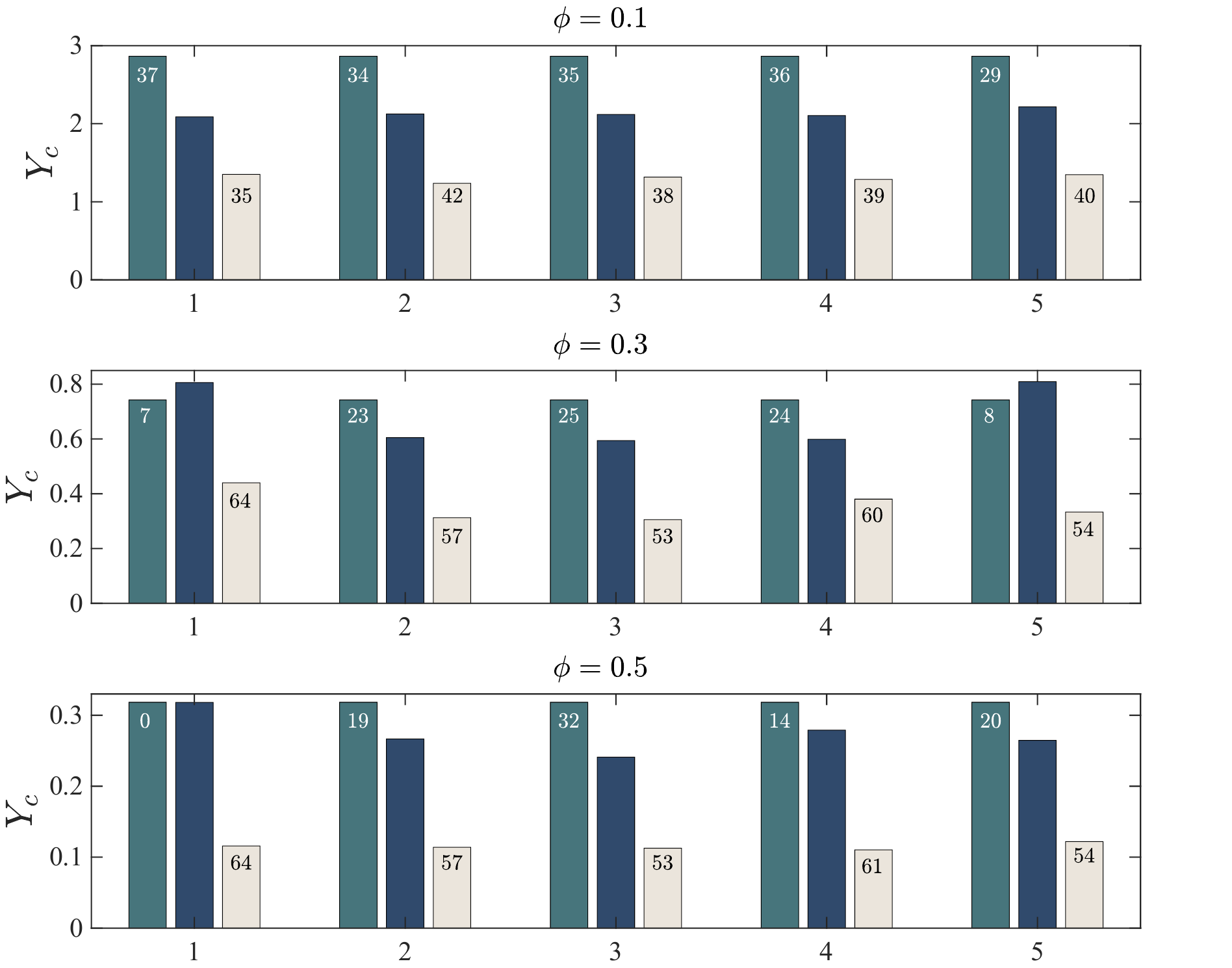}
\caption{Critical yield number ($Y_c$) for five sample realisations at different solid volume fractions, $\phi=0.1,~0.3 ~\&~ 0.5$. For each realisation, the left bar (green), the middle bar (blue) and the right bar (beige) correspond to Eq.~(\ref{eq:emad}), the DNS result, and the upper bound (\ref{eq:castaneda_yield}), respectively. The numerical values on each bar indicate the percentage error relative to the DNS result, i.e.~the absolute difference between the predicted $Y_c$ by expression (\ref{eq:emad}) or (\ref{eq:castaneda_yield}) and the DNS result divided by the DNS value.}
\label{fig:Yc}
\end{center}
\end{figure}

It should be noted that calculating $Y_c$ using the above inequality (or equality in the case of Al-Fariss \& Pinder's proposed model) requires prior knowledge of the permeability of the medium in the Newtonian limit, $\kappa_0$. In our previous work \cite{chaparian2024yielding}, to compare this bound with our model and computational data, we estimated $\kappa_0$ using the BCK model (i.e.~expression (\ref{eq:BCK_scaling})) and substituted it into expression (\ref{eq:castaneda_yield}) (see the blue line in figure 6 of \cite{chaparian2024yielding}). However, in the present study, for further comparison, we individually investigate some sample  realisations: we compute $\kappa_0$ for these sample realisations directly through numerical simulations at $B=0$, and then use this to calculate the bound for $Y_c$ by the above expression. Then we individually compare the performance of Casta{\~n}eda's bound/Al-Fariss \& Pinder's model against our model (\ref{eq:emad}) for each realisation. Fig.~\ref{fig:Yc} shows this comparison for five individual realisations chosen from three different solid ``volume'' fractions (i.e.~fifteen realisations in total).

%\begin{equation}
%Y_c = \lim_{B \to \infty} \frac{B}{\sqrt{\frac{1-\phi}{\kappa_0}} B} = \sqrt{\frac{\kappa_0}{1-\phi}} = {\color{red} \sqrt{\frac{L_{inl}/L}{(1-\phi)  (\Delta P/L)_0}} }  
%\end{equation}

Please note that our model (\ref{eq:emad}) is derived from statistical data of a wide range of solid volume fractions, and therefore predicts $Y_c$ as a function of $\phi$ alone. In other words, in our model, the critical yield number remains constant across different realisations with the same $\phi$. In Fig.~\ref{fig:Yc}, we present results for five representative realisations at $\phi = 0.1$, $0.3$ and $0.5$. For each case, the left bar (green) represents our model (\ref{eq:emad}), the middle bar (blue) corresponds to the computed data using direct numerical simulations (DNS) and the right bar (beige) is the Casta{\~n}eda's bound (\ref{eq:castaneda_yield}). The value displayed on each bar indicates the relative error compared to the DNS result, i.e.~the absolute difference between the predicted $Y_c$ by expression (\ref{eq:emad}) or (\ref{eq:castaneda_yield}) and the DNS result divided by the DNS value.

As is clear in Fig.~\ref{fig:Yc}, our model consistently outperforms the other two bound/model, particularly at higher solid volume fractions. Nonetheless, expression (\ref{eq:h_L}) can be used directly for any individual realisation. For instance, in realisation no.~5 at $\phi = 0.1$, the DNS yields $Y_c = 2.22$, expression (\ref{eq:h_L}) predicts $Y_c = 1.95$ (error $\approx$ 12.1\%), while expression (\ref{eq:emad}) gives $Y_c = 2.86$ (error $\approx$ 28.8\%). Despite this, our both predictions are notably closer to the DNS value compared to the bound (\ref{eq:castaneda_yield}) or Al-Fariss \& Pinder's model, which in this case shows an error of approximately 40\%.

\section{Darcy-type law for viscoplastic fluids} \label{sec:darcy}

The aim of this section is to propose a general Darcy-type law that predicts the pressure gradient of a viscoplastic fluid flow through porous medium as a function of its porosity and the Bingham number:
\begin{equation}
\Delta P/L = f(\phi,B),
\end{equation}
which remains reasonably accurate across all flow regimes discussed earlier. As previously outlined, when $B \to 0$, the problem approaches the Newtonian limit and the fluid flows nearly everywhere in the domain, while for $B \to \infty$, the flow is highly localised to a single path and most of the fluid remains unyielded. For this latter regime, we have already developed a model for the critical pressure gradient in the previous section,
\begin{equation}\label{eq:lim_yield}
\lim_{B \to \infty} f(\phi,B) = \pi \frac{\phi}{1-\phi}B.
\end{equation}

Now, we consider the other limit $B \to 0$. In section \ref{sec:viscous}, we reviewed several models proposed for this Newtonian limit. Among them, series-based solutions can be readily fitted to 2D numerical data with a general form,
\begin{equation}\label{eq:Newtonian_limit}
\lim_{B \to 0} f(\phi,B) = \frac{a}{-\log \phi + b + c \phi + d \phi^2 + O (\phi^3)},
\end{equation}
where compared to expression (\ref{eq:series_drag}), the unkown coefficients account for differences in drag due to obstacle shape (e.g.~square vs.~cylindrical) and the randomised distribution used in the present study, as opposed to fixed periodic arrangements (e.g.~square or regular face-to-face, hexagonal lattices, etc.).

\begin{figure}
\begin{center}
\includegraphics[width=0.6\textwidth]{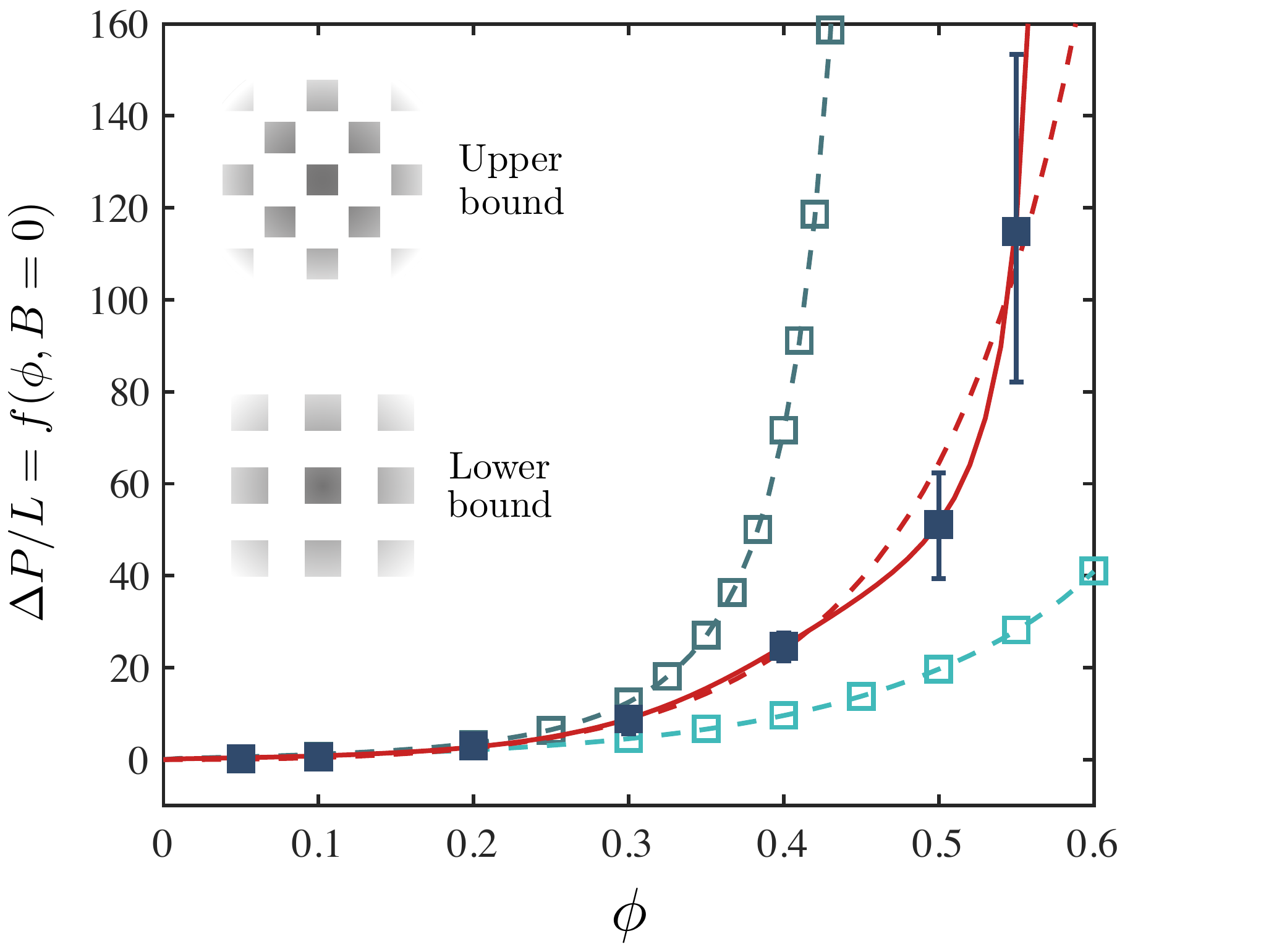}
\caption{Non-dimensional pressure gradient at $B=0$ (Newtonian case) as a function of solid ``volume'' fraction. The solid symbols with uncertainty bars represent the ensemble average of the computations for the randomised realisations; the red continuous and dashed curves are the fitted expression (\ref{eq:Newtonian_limit}) and the BCK model (i.e.~expression (\ref{eq:BCK_scaling})), respectively. The upper and lower bounds are also gathered from the computations with the triangular (i.e.~staggered) and the square (i.e.~regular face-to-face) arrangement of the obstacles, respectively, which are shown schematically in the inset. The green dashed curves are the fitted expression (\ref{eq:Newtonian_limit}) on the upper and lower bounds.}
\label{fig:Newtonian}
\end{center}
\end{figure}

Figure \ref{fig:Newtonian} presents the ensemble average of the pressure gradient versus the solid volume fraction obtained from all simulations (20 for each $\phi$) performed at $B=0$ with the randomised geometries (solid symbols with uncertainty bars). The lower and upper bounds corresponding to the square and triangular obstacle arrangements are shown as hollow symbols in lighter and darker green, respectively. The continuous red curve and the two green dashed lines are fits of expression (\ref{eq:Newtonian_limit}) into the ensemble average and the lower \& upper bounds. The fitted coefficients are reported in Table \ref{tab:table_fitting} of Appendix \ref{sec:data}. The red dashed line shows the BCK model (Eq.~(\ref{eq:BCK_scaling})) fitted with a pre-factor of 0.031. Over the range of $\phi$ investigated here, the expression (\ref{eq:Newtonian_limit}) offers a more accurate fit to our DNS data in comparison to the BCK model: R-squared is 1 for expression (\ref{eq:Newtonian_limit}) while for BCK, it is almost 0.97. Note that the BCK model is originally derived for 3D systems, although it has been previously used to interpret 2D data as well, e.g.~in \cite{shahsavari2016mobility}.

\begin{figure}
%\begin{center}
\includegraphics[width=\textwidth]{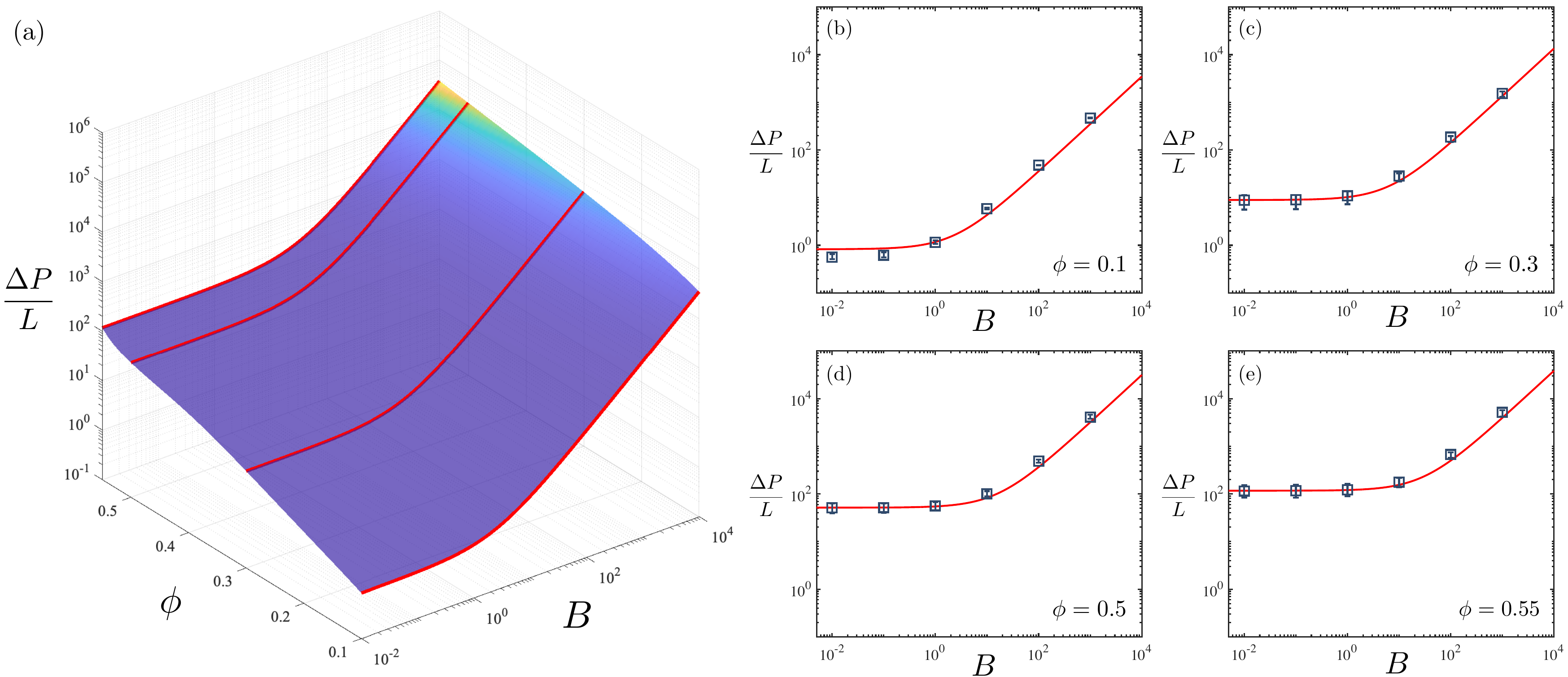}
\caption{Proposed Darcy-type law: (a) surface plot of Eq.~(\ref{eq:proposed_Darcy}) across a range of Bingham numbers and solid ``volume'' fractions. Panels (b–e) show specific cross sections of this surface at (b) $\phi=0.1$, (c) $\phi=0.3$, (d) $\phi=0.5$, and (e) $\phi=0.55$ with continuous lines. In panels (b-e), the symbols represent the ensemble average of the DNS data (randomised porous media) with uncertainty bars.}
\label{fig:Darcy}
%\end{center}
\end{figure}

For a general Darcy-type law, we now propose a simple linear combination of the Newtonian Darcy law (\ref{eq:Newtonian_limit}) with the yield limit model (\ref{eq:lim_yield}):
\begin{equation}\label{eq:proposed_Darcy}
\Delta P/L ~(\phi,B) = \lim_{B \to 0} f + \lim_{B \to \infty} f = \frac{a}{-\log \phi + b + c \phi + d \phi^2 + O (\phi^3)} + \pi \frac{\phi}{1-\phi} B,
\end{equation}
in which the free parameters in the Newtonian Darcy component (first term on the right-hand side) are determined as discussed above. Fig.~\ref{fig:Darcy}(a) illustrates the proposed expression (\ref{eq:proposed_Darcy}) as a surface in the space $B$ and $\phi$. As expected, the pressure gradient increases monotonically with both the Bingham number and the solid ``volume'' fraction. Panels (b-e) compare the proposed law (continuous curve) with the ensemble average DNS data (20 realisations for each $\phi$) for various solid ``volume'' fractions (i.e.~$\phi=0.1, 0.3, 0.5$ and $0.55$) represented by symbols and uncertainty bars. Great agreement can be evidenced with the computational data which confirms the validity of the proposed law (\ref{eq:proposed_Darcy}).

%\begin{landscape}
%\end{landscape}

\section{Summary \& conclusion}
Adaptive finite element simulations based on an augmented Lagrangian scheme were performed to anticipate the study of fluid flows of yield-stress fluids in porous media. The main aim was to propose a general Darcy-type law for viscoplastic fluids that predicts the pressure gradient required to sustain a given flow rate across the entire range of relevant parameters (encapsulated by dimensionless Bingham number, $B$). This work presented a natural extension of the framework developed in our previous study \cite{chaparian2024yielding} where a universal scale was introduced for the limiting pressure gradient that should be exceeded to initiate the flow due to the yield stress resistance. This yield limit is mimicked by $B \to \infty$ in our ``resistance formulation''. To complement this, we first adopted a model for the other missing part of the Darcy-type law puzzle --- the viscous limit, $B \to 0$ --- from the pool of previously proposed Newtonian Darcy laws in the literature, briefly reviewed in section \ref{sec:viscous}. By fitting this model to our 2D computational data (section \ref{sec:darcy}), we not only validated its applicability but also determined the corresponding fitting coefficients for randomised square-obstacle topologies. Finally, we combined the models for these two limits to derive a general Darcy-type law applicable across the full spectrum of Bingham numbers (spanning from $B \to 0$ or the viscous limit to $B \to \infty$ or the yield limit) and a wide range of solid ``volume'' fractions, namely $\phi \in [0.1,0.55]$.

Moreover, in section \ref{sec:yield}, we conducted a detailed comparison between our previously proposed model for the yield limit (i.e.~expression (\ref{eq:emad})) and the upper bound derived by Casta{\~n}eda \cite{castaneda2023variational} using a homogenisation approach (i.e.~expression (\ref{eq:castaneda_yield})). Our analysis demonstrated that our model provides a more accurate prediction of the critical yield number (inverse of the dimensionless limiting pressure gradient), particularly at higher solid ``volume'' fractions or indeed lower porosities.

Regarding our proposed Darcy-type law (i.e.~expression (\ref{eq:proposed_Darcy})), it should be noted that this expression is derived based on a combination of theoretical modelling and analysing exhaustive computational data. However, this should not be confused with the fact that individual realisations/topologies may exhibit distinct microscale characteristics. Rather, our proposed Darcy-type law is indeed a statistically averaged model that inherently filters out such microstructural variations, providing a generalised predictive framework at the macroscale.

%Notably, the proposed Darcy-type law does not require {\it a priori} knowledge of the permeability of the medium in the case of a viscous fluid flow.

%\begin{figure}
%\begin{center}
%\includegraphics[width=.7\textwidth]{Conclusion_figure.pdf}
%\caption{Bulk pressure gradient versus Bingham number for two different realisations, i.e.~no.~3 \& 4 at $\phi=0.5$ in figure \ref{fig:Yc}.}
%\label{fig:conclusion}
%\end{center}
%\end{figure}

From an application perspective, we may make a few comments regarding our model. First, the permeability of viscoplastic fluid flows is a function of rheological parameters and structure of the medium. In the case of Bingham fluids, for example, this may be expressed as,
\begin{equation}
\hat{\kappa} = \hat{f}(\hat{\mu}_p,\hat{\tau}_y,\phi).
\end{equation}
The rheological properties of the material are encapsulated in the dimensionless Bingham number in the present study. The proposed framework can be readily extended to accommodate other ``simple'' yield-stress fluid models. For example, the Herschel–Bulkley model could be incorporated by modifying the viscous limit term, $\lim_{B \to 0} f$, in expression (\ref{eq:proposed_Darcy}) to reflect shear-thinning behaviour. The yield limit term, however, remains unchanged, as the critical yield number is independent of the fluid's viscosity: in the yield limit regime, the viscous dissipation $a(\boldsymbol{u}, \boldsymbol{u})$ is at least one order of magnitude smaller than the plastic dissipation $B j(\boldsymbol{u})$. Incorporating more complex rheological behaviours such as elastoviscoplasticity \cite{chaparian2019porous,parvar2024general} into this framework is more challenging and will be the focus of our future works. Furthermore, various porous topologies should be investigated in our future studies, including polydisperse obstacles in order to expand the applicability of our proposed model.

It should be noted that, in the present study, our focus was exclusively on 2D flows. Three-dimensional analysis is left for future investigations. Nevertheless, the presented framework would also be valid in 3D, provided that suitable models for the viscous and the yield limits are combined. Although 3D viscous models are already well-established (several of those have been reviewed in the present work \S\ref{sec:viscous}), a sound yield limit mathematical model in 3D (i.e.~three-dimensional version of expression (\ref{eq:emad})) is still required. Accurate 3D DNS results or experimental data are also essential for validation purposes.

\section*{Acknowledgment}

The author appreciates Sir David Anderson Bequest Award. We also thank William Dempster for various discussions during the course of this work.

\section*{Data availability}
The data supporting this study's findings are available within the article.

\appendix
%\begin{appendix}
\section{Poiseuille flow of a Bingham fluid in a capillary tube}\label{sec:appA}
Here, we derive the velocity profile and the corresponding flow rate expressions for a Poiseuille flow of a Bingham fluid in a capillary tube with radius $\hat{R}_c$. The momentum balance equation in this case reduces to,
\begin{equation}
\frac{\text{d} \hat{p}}{\text{d} \hat{z}} = \frac{1}{\hat{r}} \frac{\text{d}}{\text{d} \hat{r}} (\hat{r} ~\hat{\tau}), ~~\text{or} ~~ \hat{\tau}=\frac{\text{d} \hat{p}}{\text{d} \hat{z}} \frac{\hat{r}}{2},
\end{equation}
where $\hat{\tau}$ is the shear stress and $z$-direction is aligned with the tube's axis. In this case, the Bingham constitutive equation (\ref{eq:const_dimensional}) reduces to,
\begin{equation}
  \left\{
    \begin{array}{ll}
      \displaystyle\frac{\text{d} \hat{p}}{\text{d} \hat{z}} \frac{\hat{r}}{2} = \hat{\mu}_p \frac{\text{d} \hat{u}}{\text{d} \hat{r}} - \hat{\tau}_y, & \mbox{iff}\quad \hat{r} > \hat{R}_Y, \\[12pt]
      \displaystyle\frac{\text{d} \hat{u}}{\text{d} \hat{r}}=0, & \mbox{iff}\quad \hat{r} \leqslant \hat{R}_Y,
  \end{array} \right.
\end{equation}
where $\hat{R}_Y$ denotes the radius of the core unyielded region or $\hat{R}_Y = -2 \frac{\hat{\tau}_y}{\text{d}p/\text{d}z} = 2 \frac{\hat{\tau}_y}{\Delta P/L}$. Hence, the velocity profile will be,
\begin{equation}\label{eq:velocity}
  \left\{
    \begin{array}{ll}
      \hat{u} = \displaystyle \frac{-1}{4\hat{\mu}_p} \frac{\Delta\hat{P}}{\hat{L}} \left( \hat{r}^2 - \hat{R}_c^2 \right) + \frac{\hat{\tau}_y}{\hat{\mu}_p} (\hat{r}-\hat{R}_c) & \mbox{iff}\quad \hat{r} > \hat{R}_Y, \\[12pt]
      \hat{u} = \hat{U}_Y & \mbox{iff}\quad \hat{r} \leqslant \hat{R}_Y,
  \end{array} \right.
\end{equation}
given that $\hat{u}(\hat{r}=\hat{R}_c)=0$. In Eq.~(\ref{eq:velocity}), $\hat{U}_Y$ is the uniform velocity of the core unyielded region. Therefore, the mean velocity in the capillary is,
\begin{equation}\label{eq:flowrate}
\hat{U} = \displaystyle \frac{\displaystyle\int_0^{\hat{R}_c} 2\pi ~\hat{u}~\hat{r} ~\text{d}\hat{r}}{\pi \hat{R}_c^2} = - \frac{1}{\hat{R}_c^2} \int_{\hat{R}_Y}^{\hat{R}_c} \hat{r}^2 \frac{\text{d}\hat{u}}{\text{d}\hat{r}} \text{d}\hat{r} = \frac{1}{\hat{\mu}_p} \left[ \frac{1}{8} \hat{R}_c^2 \frac{\Delta \hat{P}}{\hat{L}} - \frac{1}{3} \hat{R}_c ~\hat{\tau}_y + \frac{2}{3} \frac{1}{\hat{R}_c^2} \frac{\hat{\tau}^4_y}{\left( \Delta \hat{P}/\hat{L} \right)^3} \right].
\end{equation}
As is evident from this expression, the flow rate of a viscoplastic Poiseuille flow is a non-linear function of the pressure gradient and the yield stress. Al-Fariss \& Pinder \cite{alfariss1987flow} assumed that $\hat{R}_Y/\hat{R}_c \ll 1$ in deriving the flow rate expression. This is only valid in the Newtonian/viscous limit where $\hat{\tau}_y \ll (\Delta \hat{P}/\hat{L})\hat{R}_c$. In the porous media flows, however, this assumption is typically invalid, see the bottom panels in Fig.~\ref{fig:regimes} where finite unyielded regions are clear in the flow paths between the obstacles. Under this assumption, the last term in Eq.~(\ref{eq:flowrate}) is omitted, resulting in a simplified linear form of expression (\ref{eq:AlFariss_Pinder}).

\section{Generalised BCK model for a Bingham fluid}\label{sec:appB}
We derive here a generalised BCK model for Bingham fluid flows by following the same procedure outlined in section \ref{sec:BCK}. From Eqs.~(\ref{eq:BCK_derivation}) \& (\ref{eq:flowrate}), we can conclude that,
\begin{equation}
\hat{\kappa} = \left( 1-\phi \right) \left[ \frac{1}{8} \hat{R}_h^2 - \frac{1}{3} \hat{R}_h \frac{\hat{\tau}_y}{\left( \Delta \hat{P}/\hat{L} \right)} + \frac{2}{3} \frac{1}{\hat{R}_h^2} \frac{\hat{\tau}^4_y}{\left( \Delta \hat{P}/\hat{L} \right)^4} \right],
\end{equation}
where the hydraulic radius is,
\begin{equation}
\hat{R}_h = \frac{1-\phi}{\phi} \frac{\hat{R}_c}{3}.
\end{equation}
Hence,
\begin{equation}\label{eq:BCK_YSF}
\frac{\hat{\kappa}}{\hat{R}^2} = \frac{1}{72} \frac{(1-\phi)^3}{\phi^2} - \frac{1}{9} \frac{(1-\phi)^2}{\phi} \frac{\hat{\tau}_y}{(\Delta \hat{P}/\hat{L}) \hat{R}} + \frac{18}{3} \frac{\phi^2}{(1-\phi) } \frac{\hat{\tau}_y^4}{(\Delta \hat{P}/\hat{L})^4 \hat{R}^4},
\end{equation}
or in non-dimensional ``resistance'' form,
\begin{equation}
1 = \frac{1}{72} \frac{(1-\phi)^3}{\phi^2} \frac{\Delta P}{L} - \frac{1}{9} \frac{(1-\phi)^2}{\phi} B + \frac{18}{3} \frac{\phi^2}{(1-\phi) } \frac{B^4}{(\Delta \hat{P}/\hat{L})^3 }.
\end{equation}
If we retain only the first two linear terms in expression (\ref{eq:BCK_YSF}), as done by Al-Fariss \& Pinder \cite{alfariss1987flow}, the pressure gradient can be explicitly expressed in terms of the superficial velocity and the yield stress,
\begin{equation}\label{eq:pressure_linear}
\left( \frac{\Delta \hat{P}}{\hat{L}} \right)_{\text{linearised}} = 72~ \hat{\mu}_p ~\hat{U}_D \frac{1}{\hat{R}^2} \frac{\phi^2}{(1-\phi)^3} + \frac{8}{\hat{R}} \frac{\phi}{1-\phi} \hat{\tau}_y.
\end{equation}
Now, it can be alternatively explained why the yield limit term in expressions (\ref{eq:AlFariss_Pinder}) \& (\ref{eq:castaneda}) takes the form $\sqrt{(1-\phi)/\hat{\kappa}_0}$. Considering $\hat{\kappa}_0 = \displaystyle \frac{\hat{R}^2}{72} \frac{(1-\phi)^3}{\phi^2}$ (the first term in (\ref{eq:BCK_YSF})) which is indeed the BCK model for a viscous fluid, we can re-write the pre-factor of $\hat{\tau}_y$ in expression (\ref{eq:pressure_linear}) as,
\begin{equation}
\frac{8}{\hat{R}} \frac{\phi}{1-\phi} \sim  \sqrt{\frac{\phi^2}{\hat{R}^2(1-\phi)^2}} \sim \sqrt{\frac{1-\phi}{\hat{\kappa}_0}}.
\end{equation}

\section{Fitting parameters}\label{sec:data}

Table \ref{tab:table_fitting} reports the fitted parameters used in Fig.~\ref{fig:Newtonian} for the average randomised DNS data, triangular arrangement (upper bound), and square arrangement (lower bound).

\begin{table}[h]
%\begin{ruledtabular}
\begin{tabular}{ c c c c}
\hline
\hline
 &~~
\textrm{DNS data} ~&~
\textrm{Lower bound} ~&~
\textrm{Upper bound}\\
\colrule
$a$ & $0.4091$ & $0.6824$ & $0.6656$\\
$b$ & $-2.1954$ & $-1.8148$ & $-2.0637$\\
$c$ & $4.1141$ & $2.8705$ & $3.4974$\\
$d$ & $-2.1874$ & $-1.1157$ & $-1.5139$\\
\hline
\hline
\end{tabular}
%\end{ruledtabular}
\caption{\label{tab:table_fitting}%
Fitted parameters of Eq.~(\ref{eq:Newtonian_limit}) used in Fig.~\ref{fig:Newtonian}.}
\end{table}

%\end{appendix}

\bibliography{Viscoplastic}

\begin{thebibliography}{48}%
\makeatletter
\providecommand \@ifxundefined [1]{%
 \@ifx{#1\undefined}
}%
\providecommand \@ifnum [1]{%
 \ifnum #1\expandafter \@firstoftwo
 \else \expandafter \@secondoftwo
 \fi
}%
\providecommand \@ifx [1]{%
 \ifx #1\expandafter \@firstoftwo
 \else \expandafter \@secondoftwo
 \fi
}%
\providecommand \natexlab [1]{#1}%
\providecommand \enquote  [1]{``#1''}%
\providecommand \bibnamefont  [1]{#1}%
\providecommand \bibfnamefont [1]{#1}%
\providecommand \citenamefont [1]{#1}%
\providecommand \href@noop [0]{\@secondoftwo}%
\providecommand \href [0]{\begingroup \@sanitize@url \@href}%
\providecommand \@href[1]{\@@startlink{#1}\@@href}%
\providecommand \@@href[1]{\endgroup#1\@@endlink}%
\providecommand \@sanitize@url [0]{\catcode `\\12\catcode `\$12\catcode `\&12\catcode `\#12\catcode `\^12\catcode `\_12\catcode `\%12\relax}%
\providecommand \@@startlink[1]{}%
\providecommand \@@endlink[0]{}%
\providecommand \url  [0]{\begingroup\@sanitize@url \@url }%
\providecommand \@url [1]{\endgroup\@href {#1}{\urlprefix }}%
\providecommand \urlprefix  [0]{URL }%
\providecommand \Eprint [0]{\href }%
\providecommand \doibase [0]{https://doi.org/}%
\providecommand \selectlanguage [0]{\@gobble}%
\providecommand \bibinfo  [0]{\@secondoftwo}%
\providecommand \bibfield  [0]{\@secondoftwo}%
\providecommand \translation [1]{[#1]}%
\providecommand \BibitemOpen [0]{}%
\providecommand \bibitemStop [0]{}%
\providecommand \bibitemNoStop [0]{.\EOS\space}%
\providecommand \EOS [0]{\spacefactor3000\relax}%
\providecommand \BibitemShut  [1]{\csname bibitem#1\endcsname}%
\let\auto@bib@innerbib\@empty
%</preamble>
\bibitem [{\citenamefont {Darcy}(1856)}]{darcy1856fontaines}%
  \BibitemOpen
  \bibfield  {author} {\bibinfo {author} {\bibfnamefont {H.}~\bibnamefont {Darcy}},\ }\href@noop {} {\emph {\bibinfo {title} {Les fontaines publiques de la ville de Dijon}}}\ (\bibinfo  {publisher} {Dalmont},\ \bibinfo {year} {1856})\BibitemShut {NoStop}%
\bibitem [{\citenamefont {Whitaker}(1986)}]{whitaker1986flow}%
  \BibitemOpen
  \bibfield  {author} {\bibinfo {author} {\bibfnamefont {S.}~\bibnamefont {Whitaker}},\ }\bibfield  {title} {\bibinfo {title} {Flow in porous media i: A theoretical derivation of {D}arcy's law},\ }\href@noop {} {\bibfield  {journal} {\bibinfo  {journal} {Transp. Porous Med.}\ }\textbf {\bibinfo {volume} {1}},\ \bibinfo {pages} {3} (\bibinfo {year} {1986})}\BibitemShut {NoStop}%
\bibitem [{\citenamefont {Wang}\ \emph {et~al.}(2023)\citenamefont {Wang}, \citenamefont {Han}, \citenamefont {Li}, \citenamefont {Liu}, \citenamefont {Wang}, \citenamefont {Huang}, \citenamefont {Wang}, \citenamefont {Zhang},\ and\ \citenamefont {Lin}}]{wang2023review}%
  \BibitemOpen
  \bibfield  {author} {\bibinfo {author} {\bibfnamefont {Y.}~\bibnamefont {Wang}}, \bibinfo {author} {\bibfnamefont {X.}~\bibnamefont {Han}}, \bibinfo {author} {\bibfnamefont {J.}~\bibnamefont {Li}}, \bibinfo {author} {\bibfnamefont {R.}~\bibnamefont {Liu}}, \bibinfo {author} {\bibfnamefont {Q.}~\bibnamefont {Wang}}, \bibinfo {author} {\bibfnamefont {C.}~\bibnamefont {Huang}}, \bibinfo {author} {\bibfnamefont {X.}~\bibnamefont {Wang}}, \bibinfo {author} {\bibfnamefont {L.}~\bibnamefont {Zhang}},\ and\ \bibinfo {author} {\bibfnamefont {R.}~\bibnamefont {Lin}},\ }\bibfield  {title} {\bibinfo {title} {Review on oil displacement technologies of enhanced oil recovery: state-of-the-art and outlook},\ }\href@noop {} {\bibfield  {journal} {\bibinfo  {journal} {Energy \& Fuels}\ }\textbf {\bibinfo {volume} {37}},\ \bibinfo {pages} {2539} (\bibinfo {year} {2023})}\BibitemShut {NoStop}%
\bibitem [{\citenamefont {Zou}\ \emph {et~al.}(2021)\citenamefont {Zou}, \citenamefont {H\r{a}kansson},\ and\ \citenamefont {Cvetkovic}}]{Ulf2021}%
  \BibitemOpen
  \bibfield  {author} {\bibinfo {author} {\bibfnamefont {L.}~\bibnamefont {Zou}}, \bibinfo {author} {\bibfnamefont {U.}~\bibnamefont {H\r{a}kansson}},\ and\ \bibinfo {author} {\bibfnamefont {V.}~\bibnamefont {Cvetkovic}},\ }\href@noop {} {\emph {\bibinfo {title} {Analysis of cement grout propagation in fractured rocks}}},\ \bibinfo {type} {Tech. Rep.}\ (\bibinfo  {institution} {Rock Engineering Research Foundation Report 200},\ \bibinfo {year} {2021})\BibitemShut {NoStop}%
\bibitem [{\citenamefont {da~Rocha~Gomes}\ \emph {et~al.}(2023)\citenamefont {da~Rocha~Gomes}, \citenamefont {Ferrara}, \citenamefont {S{\'a}nchez},\ and\ \citenamefont {Moreno}}]{da2023comprehensive}%
  \BibitemOpen
  \bibfield  {author} {\bibinfo {author} {\bibfnamefont {S.}~\bibnamefont {da~Rocha~Gomes}}, \bibinfo {author} {\bibfnamefont {L.}~\bibnamefont {Ferrara}}, \bibinfo {author} {\bibfnamefont {L.}~\bibnamefont {S{\'a}nchez}},\ and\ \bibinfo {author} {\bibfnamefont {M.~S.}\ \bibnamefont {Moreno}},\ }\bibfield  {title} {\bibinfo {title} {A comprehensive review of cementitious grouts: Composition, properties, requirements and advanced performance},\ }\href@noop {} {\bibfield  {journal} {\bibinfo  {journal} {Constr. Build. Mater.}\ }\textbf {\bibinfo {volume} {375}},\ \bibinfo {pages} {130991} (\bibinfo {year} {2023})}\BibitemShut {NoStop}%
\bibitem [{\citenamefont {Izadi}\ \emph {et~al.}(2023)\citenamefont {Izadi}, \citenamefont {Chaparian}, \citenamefont {Trudel},\ and\ \citenamefont {Frigaard}}]{izadi2023}%
  \BibitemOpen
  \bibfield  {author} {\bibinfo {author} {\bibfnamefont {M.}~\bibnamefont {Izadi}}, \bibinfo {author} {\bibfnamefont {E.}~\bibnamefont {Chaparian}}, \bibinfo {author} {\bibfnamefont {E.}~\bibnamefont {Trudel}},\ and\ \bibinfo {author} {\bibfnamefont {I.}~\bibnamefont {Frigaard}},\ }\bibfield  {title} {\bibinfo {title} {Squeeze cementing of micro-annuli: a visco-plastic invasion flow},\ }\href@noop {} {\bibfield  {journal} {\bibinfo  {journal} {J. Non-Newtonian Fluid Mech.}\ }\textbf {\bibinfo {volume} {319}},\ \bibinfo {pages} {105070} (\bibinfo {year} {2023})}\BibitemShut {NoStop}%
\bibitem [{\citenamefont {Trivedi}\ \emph {et~al.}(2023)\citenamefont {Trivedi}, \citenamefont {Gehweiler}, \citenamefont {Wychowaniec}, \citenamefont {Ricken}, \citenamefont {Gueorguiev}, \citenamefont {Wagner},\ and\ \citenamefont {R{\"o}hrle}}]{trivedi2023continuum}%
  \BibitemOpen
  \bibfield  {author} {\bibinfo {author} {\bibfnamefont {Z.}~\bibnamefont {Trivedi}}, \bibinfo {author} {\bibfnamefont {D.}~\bibnamefont {Gehweiler}}, \bibinfo {author} {\bibfnamefont {J.~K.}\ \bibnamefont {Wychowaniec}}, \bibinfo {author} {\bibfnamefont {T.}~\bibnamefont {Ricken}}, \bibinfo {author} {\bibfnamefont {B.}~\bibnamefont {Gueorguiev}}, \bibinfo {author} {\bibfnamefont {A.}~\bibnamefont {Wagner}},\ and\ \bibinfo {author} {\bibfnamefont {O.}~\bibnamefont {R{\"o}hrle}},\ }\bibfield  {title} {\bibinfo {title} {A continuum mechanical porous media model for vertebroplasty: Numerical simulations and experimental validation},\ }\href@noop {} {\bibfield  {journal} {\bibinfo  {journal} {Biomech. Model. Mechanobiol.}\ }\textbf {\bibinfo {volume} {22}},\ \bibinfo {pages} {1253} (\bibinfo {year} {2023})}\BibitemShut {NoStop}%
\bibitem [{\citenamefont {Talon}\ and\ \citenamefont {Bauer}(2013)}]{talon2013determination}%
  \BibitemOpen
  \bibfield  {author} {\bibinfo {author} {\bibfnamefont {L.}~\bibnamefont {Talon}}\ and\ \bibinfo {author} {\bibfnamefont {D.}~\bibnamefont {Bauer}},\ }\bibfield  {title} {\bibinfo {title} {On the determination of a generalized {D}arcy equation for yield-stress fluid in porous media using a {L}attice-{B}oltzmann {TRT} scheme},\ }\href@noop {} {\bibfield  {journal} {\bibinfo  {journal} {Eur. Phys. J. E}\ }\textbf {\bibinfo {volume} {36}},\ \bibinfo {pages} {139} (\bibinfo {year} {2013})}\BibitemShut {NoStop}%
\bibitem [{\citenamefont {Liu}\ \emph {et~al.}(2019)\citenamefont {Liu}, \citenamefont {De~Luca}, \citenamefont {Rosso},\ and\ \citenamefont {Talon}}]{liu2019darcy}%
  \BibitemOpen
  \bibfield  {author} {\bibinfo {author} {\bibfnamefont {C.}~\bibnamefont {Liu}}, \bibinfo {author} {\bibfnamefont {A.}~\bibnamefont {De~Luca}}, \bibinfo {author} {\bibfnamefont {A.}~\bibnamefont {Rosso}},\ and\ \bibinfo {author} {\bibfnamefont {L.}~\bibnamefont {Talon}},\ }\bibfield  {title} {\bibinfo {title} {Darcy's law for yield stress fluids},\ }\href@noop {} {\bibfield  {journal} {\bibinfo  {journal} {Phys. Rev. Lett.}\ }\textbf {\bibinfo {volume} {122}},\ \bibinfo {pages} {245502} (\bibinfo {year} {2019})}\BibitemShut {NoStop}%
\bibitem [{\citenamefont {Chaparian}\ and\ \citenamefont {Tammisola}(2021)}]{chaparian2021sliding}%
  \BibitemOpen
  \bibfield  {author} {\bibinfo {author} {\bibfnamefont {E.}~\bibnamefont {Chaparian}}\ and\ \bibinfo {author} {\bibfnamefont {O.}~\bibnamefont {Tammisola}},\ }\bibfield  {title} {\bibinfo {title} {Sliding flows of yield-stress fluids},\ }\href@noop {} {\bibfield  {journal} {\bibinfo  {journal} {J. Fluid Mech.}\ }\textbf {\bibinfo {volume} {911}},\ \bibinfo {pages} {A17} (\bibinfo {year} {2021})}\BibitemShut {NoStop}%
\bibitem [{\citenamefont {Fraggedakis}\ \emph {et~al.}(2021)\citenamefont {Fraggedakis}, \citenamefont {Chaparian},\ and\ \citenamefont {Tammisola}}]{fraggedakis2021first}%
  \BibitemOpen
  \bibfield  {author} {\bibinfo {author} {\bibfnamefont {D.}~\bibnamefont {Fraggedakis}}, \bibinfo {author} {\bibfnamefont {E.}~\bibnamefont {Chaparian}},\ and\ \bibinfo {author} {\bibfnamefont {O.}~\bibnamefont {Tammisola}},\ }\bibfield  {title} {\bibinfo {title} {The first open channel for yield-stress fluids in porous media},\ }\href@noop {} {\bibfield  {journal} {\bibinfo  {journal} {J. Fluid Mech.}\ }\textbf {\bibinfo {volume} {911}},\ \bibinfo {pages} {A58} (\bibinfo {year} {2021})}\BibitemShut {NoStop}%
\bibitem [{\citenamefont {Chaparian}(2024)}]{chaparian2024yielding}%
  \BibitemOpen
  \bibfield  {author} {\bibinfo {author} {\bibfnamefont {E.}~\bibnamefont {Chaparian}},\ }\bibfield  {title} {\bibinfo {title} {Yielding to percolation: a universal scale},\ }\href@noop {} {\bibfield  {journal} {\bibinfo  {journal} {J. Fluid Mech.}\ }\textbf {\bibinfo {volume} {980}},\ \bibinfo {pages} {A14} (\bibinfo {year} {2024})}\BibitemShut {NoStop}%
\bibitem [{\citenamefont {Frigaard}\ \emph {et~al.}(2017)\citenamefont {Frigaard}, \citenamefont {Paso},\ and\ \citenamefont {de~Souza~Mendes}}]{frigaard2017bingham}%
  \BibitemOpen
  \bibfield  {author} {\bibinfo {author} {\bibfnamefont {I.~A.}\ \bibnamefont {Frigaard}}, \bibinfo {author} {\bibfnamefont {K.~G.}\ \bibnamefont {Paso}},\ and\ \bibinfo {author} {\bibfnamefont {P.~R.}\ \bibnamefont {de~Souza~Mendes}},\ }\bibfield  {title} {\bibinfo {title} {{B}ingham’s model in the oil and gas industry},\ }\href@noop {} {\bibfield  {journal} {\bibinfo  {journal} {Rheol. Acta}\ }\textbf {\bibinfo {volume} {56}},\ \bibinfo {pages} {259} (\bibinfo {year} {2017})}\BibitemShut {NoStop}%
\bibitem [{\citenamefont {Waisbord}\ \emph {et~al.}(2019)\citenamefont {Waisbord}, \citenamefont {Stoop}, \citenamefont {Walkama}, \citenamefont {Dunkel},\ and\ \citenamefont {Guasto}}]{waisbord2019anomalous}%
  \BibitemOpen
  \bibfield  {author} {\bibinfo {author} {\bibfnamefont {N.}~\bibnamefont {Waisbord}}, \bibinfo {author} {\bibfnamefont {N.}~\bibnamefont {Stoop}}, \bibinfo {author} {\bibfnamefont {D.~M.}\ \bibnamefont {Walkama}}, \bibinfo {author} {\bibfnamefont {J.}~\bibnamefont {Dunkel}},\ and\ \bibinfo {author} {\bibfnamefont {J.~S.}\ \bibnamefont {Guasto}},\ }\bibfield  {title} {\bibinfo {title} {Anomalous percolation flow transition of yield stress fluids in porous media},\ }\href@noop {} {\bibfield  {journal} {\bibinfo  {journal} {Phys. Rev. Fluids}\ }\textbf {\bibinfo {volume} {4}},\ \bibinfo {pages} {063303} (\bibinfo {year} {2019})}\BibitemShut {NoStop}%
\bibitem [{\citenamefont {Chaparian}\ \emph {et~al.}(2020)\citenamefont {Chaparian}, \citenamefont {Izbassarov}, \citenamefont {De~Vita}, \citenamefont {Brandt},\ and\ \citenamefont {Tammisola}}]{chaparian2019porous}%
  \BibitemOpen
  \bibfield  {author} {\bibinfo {author} {\bibfnamefont {E.}~\bibnamefont {Chaparian}}, \bibinfo {author} {\bibfnamefont {D.}~\bibnamefont {Izbassarov}}, \bibinfo {author} {\bibfnamefont {F.}~\bibnamefont {De~Vita}}, \bibinfo {author} {\bibfnamefont {L.}~\bibnamefont {Brandt}},\ and\ \bibinfo {author} {\bibfnamefont {O.}~\bibnamefont {Tammisola}},\ }\bibfield  {title} {\bibinfo {title} {Yield-stress fluids in porous media: a comparison of viscoplastic and elastoviscoplastic flows},\ }\href@noop {} {\bibfield  {journal} {\bibinfo  {journal} {Meccanica}\ }\textbf {\bibinfo {volume} {55}},\ \bibinfo {pages} {331} (\bibinfo {year} {2020})}\BibitemShut {NoStop}%
\bibitem [{\citenamefont {Parvar}\ \emph {et~al.}(2024)\citenamefont {Parvar}, \citenamefont {Chaparian},\ and\ \citenamefont {Tammisola}}]{parvar2024general}%
  \BibitemOpen
  \bibfield  {author} {\bibinfo {author} {\bibfnamefont {S.}~\bibnamefont {Parvar}}, \bibinfo {author} {\bibfnamefont {E.}~\bibnamefont {Chaparian}},\ and\ \bibinfo {author} {\bibfnamefont {O.}~\bibnamefont {Tammisola}},\ }\bibfield  {title} {\bibinfo {title} {General hydrodynamic features of elastoviscoplastic fluid flows through randomised porous media},\ }\href@noop {} {\bibfield  {journal} {\bibinfo  {journal} {Theor. Comput. Fluid Dyn.}\ }\textbf {\bibinfo {volume} {38}},\ \bibinfo {pages} {531} (\bibinfo {year} {2024})}\BibitemShut {NoStop}%
\bibitem [{\citenamefont {Al-Fariss}\ and\ \citenamefont {Pinder}(1987)}]{alfariss1987flow}%
  \BibitemOpen
  \bibfield  {author} {\bibinfo {author} {\bibfnamefont {T.}~\bibnamefont {Al-Fariss}}\ and\ \bibinfo {author} {\bibfnamefont {K.~L.}\ \bibnamefont {Pinder}},\ }\bibfield  {title} {\bibinfo {title} {Flow through porous media of a shear-thinning liquid with yield stress},\ }\href@noop {} {\bibfield  {journal} {\bibinfo  {journal} {Can. J. Chem. Eng.}\ }\textbf {\bibinfo {volume} {65}},\ \bibinfo {pages} {391} (\bibinfo {year} {1987})}\BibitemShut {NoStop}%
\bibitem [{\citenamefont {Hewitt}\ \emph {et~al.}(2016)\citenamefont {Hewitt}, \citenamefont {Daneshi}, \citenamefont {Balmforth},\ and\ \citenamefont {Martinez}}]{hewitt2016heleshaw}%
  \BibitemOpen
  \bibfield  {author} {\bibinfo {author} {\bibfnamefont {D.~R.}\ \bibnamefont {Hewitt}}, \bibinfo {author} {\bibfnamefont {M.}~\bibnamefont {Daneshi}}, \bibinfo {author} {\bibfnamefont {N.~J.}\ \bibnamefont {Balmforth}},\ and\ \bibinfo {author} {\bibfnamefont {D.~M.}\ \bibnamefont {Martinez}},\ }\bibfield  {title} {\bibinfo {title} {Obstructed and channelized viscoplastic flow in a {H}ele-{S}haw cell},\ }\href@noop {} {\bibfield  {journal} {\bibinfo  {journal} {J. Fluid Mech.}\ }\textbf {\bibinfo {volume} {790}},\ \bibinfo {pages} {173} (\bibinfo {year} {2016})}\BibitemShut {NoStop}%
\bibitem [{\citenamefont {Daneshi}\ \emph {et~al.}(2020)\citenamefont {Daneshi}, \citenamefont {MacKenzie}, \citenamefont {Balmforth}, \citenamefont {Martinez},\ and\ \citenamefont {Hewitt}}]{daneshi2020obstructed}%
  \BibitemOpen
  \bibfield  {author} {\bibinfo {author} {\bibfnamefont {M.}~\bibnamefont {Daneshi}}, \bibinfo {author} {\bibfnamefont {J.}~\bibnamefont {MacKenzie}}, \bibinfo {author} {\bibfnamefont {N.~J.}\ \bibnamefont {Balmforth}}, \bibinfo {author} {\bibfnamefont {D.~M.}\ \bibnamefont {Martinez}},\ and\ \bibinfo {author} {\bibfnamefont {D.~R.}\ \bibnamefont {Hewitt}},\ }\bibfield  {title} {\bibinfo {title} {Obstructed viscoplastic flow in a {H}ele-{S}haw cell},\ }\href@noop {} {\bibfield  {journal} {\bibinfo  {journal} {Phys. Rev. Fluids}\ }\textbf {\bibinfo {volume} {5}},\ \bibinfo {pages} {013301} (\bibinfo {year} {2020})}\BibitemShut {NoStop}%
\bibitem [{\citenamefont {Chevalier}\ \emph {et~al.}(2013)\citenamefont {Chevalier}, \citenamefont {Chevalier}, \citenamefont {Clain}, \citenamefont {Dupla}, \citenamefont {Canou}, \citenamefont {Rodts},\ and\ \citenamefont {Coussot}}]{chevalier2013darcy}%
  \BibitemOpen
  \bibfield  {author} {\bibinfo {author} {\bibfnamefont {T.}~\bibnamefont {Chevalier}}, \bibinfo {author} {\bibfnamefont {C.}~\bibnamefont {Chevalier}}, \bibinfo {author} {\bibfnamefont {X.}~\bibnamefont {Clain}}, \bibinfo {author} {\bibfnamefont {J.~C.}\ \bibnamefont {Dupla}}, \bibinfo {author} {\bibfnamefont {J.}~\bibnamefont {Canou}}, \bibinfo {author} {\bibfnamefont {S.}~\bibnamefont {Rodts}},\ and\ \bibinfo {author} {\bibfnamefont {P.}~\bibnamefont {Coussot}},\ }\bibfield  {title} {\bibinfo {title} {Darcy's law for yield stress fluid flowing through a porous medium},\ }\href@noop {} {\bibfield  {journal} {\bibinfo  {journal} {J. Non-Newtonian Fluid Mech.}\ }\textbf {\bibinfo {volume} {195}},\ \bibinfo {pages} {57} (\bibinfo {year} {2013})}\BibitemShut {NoStop}%
\bibitem [{\citenamefont {Chevalier}\ \emph {et~al.}(2014)\citenamefont {Chevalier}, \citenamefont {Rodts}, \citenamefont {Chateau}, \citenamefont {Chevalier},\ and\ \citenamefont {Coussot}}]{chevalier2014breaking}%
  \BibitemOpen
  \bibfield  {author} {\bibinfo {author} {\bibfnamefont {T.}~\bibnamefont {Chevalier}}, \bibinfo {author} {\bibfnamefont {S.}~\bibnamefont {Rodts}}, \bibinfo {author} {\bibfnamefont {X.}~\bibnamefont {Chateau}}, \bibinfo {author} {\bibfnamefont {C.}~\bibnamefont {Chevalier}},\ and\ \bibinfo {author} {\bibfnamefont {P.}~\bibnamefont {Coussot}},\ }\bibfield  {title} {\bibinfo {title} {Breaking of non-{N}ewtonian character in flows through a porous medium},\ }\href@noop {} {\bibfield  {journal} {\bibinfo  {journal} {Phys. Rev. E}\ }\textbf {\bibinfo {volume} {89}},\ \bibinfo {pages} {023002} (\bibinfo {year} {2014})}\BibitemShut {NoStop}%
\bibitem [{\citenamefont {Bleyer}\ and\ \citenamefont {Coussot}(2014)}]{bleyer2014breakage}%
  \BibitemOpen
  \bibfield  {author} {\bibinfo {author} {\bibfnamefont {J.}~\bibnamefont {Bleyer}}\ and\ \bibinfo {author} {\bibfnamefont {P.}~\bibnamefont {Coussot}},\ }\bibfield  {title} {\bibinfo {title} {Breakage of non-{N}ewtonian character in flow through a porous medium: evidence from numerical simulation},\ }\href@noop {} {\bibfield  {journal} {\bibinfo  {journal} {Phys. Rev. E}\ }\textbf {\bibinfo {volume} {89}},\ \bibinfo {pages} {063018} (\bibinfo {year} {2014})}\BibitemShut {NoStop}%
\bibitem [{\citenamefont {Shahsavari}\ and\ \citenamefont {McKinley}(2016)}]{shahsavari2016mobility}%
  \BibitemOpen
  \bibfield  {author} {\bibinfo {author} {\bibfnamefont {S.}~\bibnamefont {Shahsavari}}\ and\ \bibinfo {author} {\bibfnamefont {G.~H.}\ \bibnamefont {McKinley}},\ }\bibfield  {title} {\bibinfo {title} {Mobility and pore-scale fluid dynamics of rate-dependent yield-stress fluids flowing through fibrous porous media},\ }\href@noop {} {\bibfield  {journal} {\bibinfo  {journal} {J. Non-Newtonian Fluid Mech.}\ }\textbf {\bibinfo {volume} {235}},\ \bibinfo {pages} {76} (\bibinfo {year} {2016})}\BibitemShut {NoStop}%
\bibitem [{\citenamefont {Shahsavari}\ and\ \citenamefont {McKinley}(2015)}]{shahsavari2015mobility}%
  \BibitemOpen
  \bibfield  {author} {\bibinfo {author} {\bibfnamefont {S.}~\bibnamefont {Shahsavari}}\ and\ \bibinfo {author} {\bibfnamefont {G.~H.}\ \bibnamefont {McKinley}},\ }\bibfield  {title} {\bibinfo {title} {Mobility of power-law and {C}arreau fluids through fibrous media},\ }\href@noop {} {\bibfield  {journal} {\bibinfo  {journal} {Phys. Rev. E}\ }\textbf {\bibinfo {volume} {92}},\ \bibinfo {pages} {063012} (\bibinfo {year} {2015})}\BibitemShut {NoStop}%
\bibitem [{\citenamefont {Chevalier}\ and\ \citenamefont {Talon}(2015)}]{chevalier2015generalization}%
  \BibitemOpen
  \bibfield  {author} {\bibinfo {author} {\bibfnamefont {T.}~\bibnamefont {Chevalier}}\ and\ \bibinfo {author} {\bibfnamefont {L.}~\bibnamefont {Talon}},\ }\bibfield  {title} {\bibinfo {title} {Generalization of {D}arcy's law for {B}ingham fluids in porous media: From flow-field statistics to the flow-rate regimes},\ }\href@noop {} {\bibfield  {journal} {\bibinfo  {journal} {Phys. Rev. E}\ }\textbf {\bibinfo {volume} {91}},\ \bibinfo {pages} {023011} (\bibinfo {year} {2015})}\BibitemShut {NoStop}%
\bibitem [{\citenamefont {Bauer}\ \emph {et~al.}(2019)\citenamefont {Bauer}, \citenamefont {Talon}, \citenamefont {Peysson}, \citenamefont {Ly}, \citenamefont {Bat{\^o}t}, \citenamefont {Chevalier},\ and\ \citenamefont {Fleury}}]{bauer2019experimental}%
  \BibitemOpen
  \bibfield  {author} {\bibinfo {author} {\bibfnamefont {D.}~\bibnamefont {Bauer}}, \bibinfo {author} {\bibfnamefont {L.}~\bibnamefont {Talon}}, \bibinfo {author} {\bibfnamefont {Y.}~\bibnamefont {Peysson}}, \bibinfo {author} {\bibfnamefont {H.~B.}\ \bibnamefont {Ly}}, \bibinfo {author} {\bibfnamefont {G.}~\bibnamefont {Bat{\^o}t}}, \bibinfo {author} {\bibfnamefont {T.}~\bibnamefont {Chevalier}},\ and\ \bibinfo {author} {\bibfnamefont {M.}~\bibnamefont {Fleury}},\ }\bibfield  {title} {\bibinfo {title} {Experimental and numerical determination of {D}arcy's law for yield stress fluids in porous media},\ }\href@noop {} {\bibfield  {journal} {\bibinfo  {journal} {Phys. Rev. Fluids}\ }\textbf {\bibinfo {volume} {4}},\ \bibinfo {pages} {063301} (\bibinfo {year} {2019})}\BibitemShut {NoStop}%
\bibitem [{\citenamefont {Casta{\~n}eda}(2023)}]{castaneda2023variational}%
  \BibitemOpen
  \bibfield  {author} {\bibinfo {author} {\bibfnamefont {P.~P.}\ \bibnamefont {Casta{\~n}eda}},\ }\bibfield  {title} {\bibinfo {title} {Variational linear comparison homogenization estimates for the flow of yield stress fluids through porous media},\ }\href@noop {} {\bibfield  {journal} {\bibinfo  {journal} {J. Non-Newtonian Fluid Mech.}\ }\textbf {\bibinfo {volume} {321}},\ \bibinfo {pages} {105104} (\bibinfo {year} {2023})}\BibitemShut {NoStop}%
\bibitem [{\citenamefont {Roquet}\ and\ \citenamefont {Saramito}(2003)}]{roquet2003adaptive}%
  \BibitemOpen
  \bibfield  {author} {\bibinfo {author} {\bibfnamefont {N.}~\bibnamefont {Roquet}}\ and\ \bibinfo {author} {\bibfnamefont {P.}~\bibnamefont {Saramito}},\ }\bibfield  {title} {\bibinfo {title} {An adaptive finite element method for {B}ingham fluid flows around a cylinder},\ }\href@noop {} {\bibfield  {journal} {\bibinfo  {journal} {Comput. Meth. Appl. Mech. Eng.}\ }\textbf {\bibinfo {volume} {192}},\ \bibinfo {pages} {3317} (\bibinfo {year} {2003})}\BibitemShut {NoStop}%
\bibitem [{\citenamefont {Chaparian}\ and\ \citenamefont {Tammisola}(2019)}]{chaparian2019adaptive}%
  \BibitemOpen
  \bibfield  {author} {\bibinfo {author} {\bibfnamefont {E.}~\bibnamefont {Chaparian}}\ and\ \bibinfo {author} {\bibfnamefont {O.}~\bibnamefont {Tammisola}},\ }\bibfield  {title} {\bibinfo {title} {An adaptive finite element method for elastoviscoplastic fluid flows},\ }\href@noop {} {\bibfield  {journal} {\bibinfo  {journal} {J. Non-Newtonian Fluid Mech.}\ }\textbf {\bibinfo {volume} {271}},\ \bibinfo {pages} {104148} (\bibinfo {year} {2019})}\BibitemShut {NoStop}%
\bibitem [{\citenamefont {Chaparian}\ \emph {et~al.}(2022)\citenamefont {Chaparian}, \citenamefont {Owens},\ and\ \citenamefont {McKinley}}]{chaparian2022vane}%
  \BibitemOpen
  \bibfield  {author} {\bibinfo {author} {\bibfnamefont {E.}~\bibnamefont {Chaparian}}, \bibinfo {author} {\bibfnamefont {C.~E.}\ \bibnamefont {Owens}},\ and\ \bibinfo {author} {\bibfnamefont {G.~H.}\ \bibnamefont {McKinley}},\ }\bibfield  {title} {\bibinfo {title} {Computational rheometry of yielding and viscoplastic flow in vane-and-cup rheometer fixtures},\ }\href@noop {} {\bibfield  {journal} {\bibinfo  {journal} {J. Non-Newtonian Fluid Mech.}\ }\textbf {\bibinfo {volume} {307}},\ \bibinfo {pages} {104857} (\bibinfo {year} {2022})}\BibitemShut {NoStop}%
\bibitem [{\citenamefont {Prager}(1961)}]{prager1}%
  \BibitemOpen
  \bibfield  {author} {\bibinfo {author} {\bibfnamefont {S.}~\bibnamefont {Prager}},\ }\bibfield  {title} {\bibinfo {title} {Viscous flow through porous media},\ }\href@noop {} {\bibfield  {journal} {\bibinfo  {journal} {Phys. Fluids}\ }\textbf {\bibinfo {volume} {4}},\ \bibinfo {pages} {1477} (\bibinfo {year} {1961})}\BibitemShut {NoStop}%
\bibitem [{\citenamefont {Weissberg}\ and\ \citenamefont {Prager}(1962)}]{prager2}%
  \BibitemOpen
  \bibfield  {author} {\bibinfo {author} {\bibfnamefont {H.~L.}\ \bibnamefont {Weissberg}}\ and\ \bibinfo {author} {\bibfnamefont {S.}~\bibnamefont {Prager}},\ }\bibfield  {title} {\bibinfo {title} {Viscous flow through porous media. ii. approximate three-point correlation function},\ }\href@noop {} {\bibfield  {journal} {\bibinfo  {journal} {Phys. Fluids}\ }\textbf {\bibinfo {volume} {5}},\ \bibinfo {pages} {1390} (\bibinfo {year} {1962})}\BibitemShut {NoStop}%
\bibitem [{\citenamefont {Weissberg}\ and\ \citenamefont {Prager}(1970)}]{prager3}%
  \BibitemOpen
  \bibfield  {author} {\bibinfo {author} {\bibfnamefont {H.~L.}\ \bibnamefont {Weissberg}}\ and\ \bibinfo {author} {\bibfnamefont {S.}~\bibnamefont {Prager}},\ }\bibfield  {title} {\bibinfo {title} {Viscous flow through porous media. iii. upper bounds on the permeability for a simple random geometry},\ }\href@noop {} {\bibfield  {journal} {\bibinfo  {journal} {Phys. Fluids}\ }\textbf {\bibinfo {volume} {13}},\ \bibinfo {pages} {2958} (\bibinfo {year} {1970})}\BibitemShut {NoStop}%
\bibitem [{\citenamefont {Doi}(1976)}]{doi1976new}%
  \BibitemOpen
  \bibfield  {author} {\bibinfo {author} {\bibfnamefont {M.}~\bibnamefont {Doi}},\ }\bibfield  {title} {\bibinfo {title} {A new variational approach to the diffusion and the flow problem in porous media},\ }\href@noop {} {\bibfield  {journal} {\bibinfo  {journal} {J. Phys. Soc. Jpn.}\ }\textbf {\bibinfo {volume} {40}},\ \bibinfo {pages} {567} (\bibinfo {year} {1976})}\BibitemShut {NoStop}%
\bibitem [{\citenamefont {Berryman}(1985)}]{berryman1985bounds}%
  \BibitemOpen
  \bibfield  {author} {\bibinfo {author} {\bibfnamefont {J.~G.}\ \bibnamefont {Berryman}},\ }\bibfield  {title} {\bibinfo {title} {Bounds on fluid permeability for viscous flow through porous media},\ }\href@noop {} {\bibfield  {journal} {\bibinfo  {journal} {J. Chem. Phys.}\ }\textbf {\bibinfo {volume} {82}},\ \bibinfo {pages} {1459} (\bibinfo {year} {1985})}\BibitemShut {NoStop}%
\bibitem [{\citenamefont {Bignonnet}(2018)}]{bignonnet2018upper}%
  \BibitemOpen
  \bibfield  {author} {\bibinfo {author} {\bibfnamefont {F.}~\bibnamefont {Bignonnet}},\ }\bibfield  {title} {\bibinfo {title} {Upper bounds on the permeability of random porous media},\ }\href@noop {} {\bibfield  {journal} {\bibinfo  {journal} {Transp. Porous Med.}\ }\textbf {\bibinfo {volume} {122}},\ \bibinfo {pages} {57} (\bibinfo {year} {2018})}\BibitemShut {NoStop}%
\bibitem [{\citenamefont {MacDonald}\ \emph {et~al.}(1991)\citenamefont {MacDonald}, \citenamefont {Chu}, \citenamefont {Guilloit},\ and\ \citenamefont {Ng}}]{macdonald1991generalized}%
  \BibitemOpen
  \bibfield  {author} {\bibinfo {author} {\bibfnamefont {M.~J.}\ \bibnamefont {MacDonald}}, \bibinfo {author} {\bibfnamefont {C.}~\bibnamefont {Chu}}, \bibinfo {author} {\bibfnamefont {P.~P.}\ \bibnamefont {Guilloit}},\ and\ \bibinfo {author} {\bibfnamefont {K.~M.}\ \bibnamefont {Ng}},\ }\bibfield  {title} {\bibinfo {title} {A generalized {B}lake-{K}ozeny equation for multisized spherical particles},\ }\href@noop {} {\bibfield  {journal} {\bibinfo  {journal} {AIChE J.}\ }\textbf {\bibinfo {volume} {37}},\ \bibinfo {pages} {1583} (\bibinfo {year} {1991})}\BibitemShut {NoStop}%
\bibitem [{\citenamefont {Guyon}\ \emph {et~al.}(2015)\citenamefont {Guyon}, \citenamefont {Hulin}, \citenamefont {Petit},\ and\ \citenamefont {Mitescu}}]{guyon2015physical}%
  \BibitemOpen
  \bibfield  {author} {\bibinfo {author} {\bibfnamefont {E.}~\bibnamefont {Guyon}}, \bibinfo {author} {\bibfnamefont {J.~P.}\ \bibnamefont {Hulin}}, \bibinfo {author} {\bibfnamefont {L.}~\bibnamefont {Petit}},\ and\ \bibinfo {author} {\bibfnamefont {C.~D.}\ \bibnamefont {Mitescu}},\ }\href@noop {} {\emph {\bibinfo {title} {Physical hydrodynamics}}}\ (\bibinfo  {publisher} {Oxford university press},\ \bibinfo {year} {2015})\BibitemShut {NoStop}%
\bibitem [{\citenamefont {Zick}\ and\ \citenamefont {Homsy}(1982)}]{zick1982stokes}%
  \BibitemOpen
  \bibfield  {author} {\bibinfo {author} {\bibfnamefont {A.~A.}\ \bibnamefont {Zick}}\ and\ \bibinfo {author} {\bibfnamefont {G.~M.}\ \bibnamefont {Homsy}},\ }\bibfield  {title} {\bibinfo {title} {Stokes flow through periodic arrays of spheres},\ }\href@noop {} {\bibfield  {journal} {\bibinfo  {journal} {J. Fluid Mech.}\ }\textbf {\bibinfo {volume} {115}},\ \bibinfo {pages} {13} (\bibinfo {year} {1982})}\BibitemShut {NoStop}%
\bibitem [{\citenamefont {Brinkman}(1947)}]{brinkman1947}%
  \BibitemOpen
  \bibfield  {author} {\bibinfo {author} {\bibfnamefont {H.~C.}\ \bibnamefont {Brinkman}},\ }\bibfield  {title} {\bibinfo {title} {A calculation of the viscous force exerted by a flowing fluid on a dense swarm of particles},\ }\href@noop {} {\bibfield  {journal} {\bibinfo  {journal} {Appl. Sci. Res.}\ }\textbf {\bibinfo {volume} {AI}},\ \bibinfo {pages} {27} (\bibinfo {year} {1947})}\BibitemShut {NoStop}%
\bibitem [{\citenamefont {Cancelliere}\ \emph {et~al.}(1990)\citenamefont {Cancelliere}, \citenamefont {Chang}, \citenamefont {Foti}, \citenamefont {Rothman},\ and\ \citenamefont {Succi}}]{cancelliere1990permeability}%
  \BibitemOpen
  \bibfield  {author} {\bibinfo {author} {\bibfnamefont {A.}~\bibnamefont {Cancelliere}}, \bibinfo {author} {\bibfnamefont {C.}~\bibnamefont {Chang}}, \bibinfo {author} {\bibfnamefont {E.}~\bibnamefont {Foti}}, \bibinfo {author} {\bibfnamefont {D.~H.}\ \bibnamefont {Rothman}},\ and\ \bibinfo {author} {\bibfnamefont {S.}~\bibnamefont {Succi}},\ }\bibfield  {title} {\bibinfo {title} {The permeability of a random medium: comparison of simulation with theory},\ }\href@noop {} {\bibfield  {journal} {\bibinfo  {journal} {Phys. Fluids A: Fluid Dyn.}\ }\textbf {\bibinfo {volume} {2}},\ \bibinfo {pages} {2085} (\bibinfo {year} {1990})}\BibitemShut {NoStop}%
\bibitem [{\citenamefont {Hasimoto}(1959)}]{hasimoto1959periodic}%
  \BibitemOpen
  \bibfield  {author} {\bibinfo {author} {\bibfnamefont {H.}~\bibnamefont {Hasimoto}},\ }\bibfield  {title} {\bibinfo {title} {On the periodic fundamental solutions of the {S}tokes equations and their application to viscous flow past a cubic array of spheres},\ }\href@noop {} {\bibfield  {journal} {\bibinfo  {journal} {J. Fluid Mech.}\ }\textbf {\bibinfo {volume} {5}},\ \bibinfo {pages} {317} (\bibinfo {year} {1959})}\BibitemShut {NoStop}%
\bibitem [{\citenamefont {Drummond}\ and\ \citenamefont {Tahir}(1984)}]{drummond1984laminar}%
  \BibitemOpen
  \bibfield  {author} {\bibinfo {author} {\bibfnamefont {J.~E.}\ \bibnamefont {Drummond}}\ and\ \bibinfo {author} {\bibfnamefont {M.~I.}\ \bibnamefont {Tahir}},\ }\bibfield  {title} {\bibinfo {title} {Laminar viscous flow through regular arrays of parallel solid cylinders},\ }\href@noop {} {\bibfield  {journal} {\bibinfo  {journal} {Int. J. Multiph. Flow}\ }\textbf {\bibinfo {volume} {10}},\ \bibinfo {pages} {515} (\bibinfo {year} {1984})}\BibitemShut {NoStop}%
\bibitem [{\citenamefont {Hecht}(2012)}]{freefem}%
  \BibitemOpen
  \bibfield  {author} {\bibinfo {author} {\bibfnamefont {F.}~\bibnamefont {Hecht}},\ }\bibfield  {title} {\bibinfo {title} {New development in freefem++},\ }\href@noop {} {\bibfield  {journal} {\bibinfo  {journal} {J. Numer. Math.}\ }\textbf {\bibinfo {volume} {20}},\ \bibinfo {pages} {251} (\bibinfo {year} {2012})}\BibitemShut {NoStop}%
\bibitem [{\citenamefont {Chaparian}\ and\ \citenamefont {Frigaard}(2017)}]{chaparian2017yield}%
  \BibitemOpen
  \bibfield  {author} {\bibinfo {author} {\bibfnamefont {E.}~\bibnamefont {Chaparian}}\ and\ \bibinfo {author} {\bibfnamefont {I.~A.}\ \bibnamefont {Frigaard}},\ }\bibfield  {title} {\bibinfo {title} {Yield limit analysis of particle motion in a yield-stress fluid},\ }\href@noop {} {\bibfield  {journal} {\bibinfo  {journal} {J. Fluid Mech.}\ }\textbf {\bibinfo {volume} {819}},\ \bibinfo {pages} {311} (\bibinfo {year} {2017})}\BibitemShut {NoStop}%
\bibitem [{\citenamefont {Iglesias}\ \emph {et~al.}(2020)\citenamefont {Iglesias}, \citenamefont {Mercier}, \citenamefont {Chaparian},\ and\ \citenamefont {Frigaard}}]{iglesias2020computing}%
  \BibitemOpen
  \bibfield  {author} {\bibinfo {author} {\bibfnamefont {J.~A.}\ \bibnamefont {Iglesias}}, \bibinfo {author} {\bibfnamefont {G.}~\bibnamefont {Mercier}}, \bibinfo {author} {\bibfnamefont {E.}~\bibnamefont {Chaparian}},\ and\ \bibinfo {author} {\bibfnamefont {I.~A.}\ \bibnamefont {Frigaard}},\ }\bibfield  {title} {\bibinfo {title} {Computing the yield limit in three-dimensional flows of a yield stress fluid about a settling particle},\ }\href@noop {} {\bibfield  {journal} {\bibinfo  {journal} {J. Non-Newtonian Fluid Mech.}\ }\textbf {\bibinfo {volume} {284}},\ \bibinfo {pages} {104374} (\bibinfo {year} {2020})}\BibitemShut {NoStop}%
\bibitem [{\citenamefont {Medina-Ba{\~n}uelos}\ \emph {et~al.}(2023)\citenamefont {Medina-Ba{\~n}uelos}, \citenamefont {Mar{\'\i}n-Santib{\'a}{\~n}ez}, \citenamefont {Chaparian}, \citenamefont {Owens}, \citenamefont {McKinley},\ and\ \citenamefont {P{\'e}rez-Gonz{\'a}lez}}]{medina2023rheo}%
  \BibitemOpen
  \bibfield  {author} {\bibinfo {author} {\bibfnamefont {E.~F.}\ \bibnamefont {Medina-Ba{\~n}uelos}}, \bibinfo {author} {\bibfnamefont {B.~M.}\ \bibnamefont {Mar{\'\i}n-Santib{\'a}{\~n}ez}}, \bibinfo {author} {\bibfnamefont {E.}~\bibnamefont {Chaparian}}, \bibinfo {author} {\bibfnamefont {C.~E.}\ \bibnamefont {Owens}}, \bibinfo {author} {\bibfnamefont {G.~H.}\ \bibnamefont {McKinley}},\ and\ \bibinfo {author} {\bibfnamefont {J.}~\bibnamefont {P{\'e}rez-Gonz{\'a}lez}},\ }\bibfield  {title} {\bibinfo {title} {Rheo-{PIV} of yield-stress fluids in a {3D}-printed fractal vane-in-cup geometry},\ }\href@noop {} {\bibfield  {journal} {\bibinfo  {journal} {J. Rheol.}\ }\textbf {\bibinfo {volume} {67}},\ \bibinfo {pages} {891} (\bibinfo {year} {2023})}\BibitemShut {NoStop}%
\bibitem [{\citenamefont {Chaparian}\ and\ \citenamefont {Tammisola}(2020)}]{chaparian2020stability}%
  \BibitemOpen
  \bibfield  {author} {\bibinfo {author} {\bibfnamefont {E.}~\bibnamefont {Chaparian}}\ and\ \bibinfo {author} {\bibfnamefont {O.}~\bibnamefont {Tammisola}},\ }\bibfield  {title} {\bibinfo {title} {Stability of particles inside yield-stress fluid {P}oiseuille flows},\ }\href@noop {} {\bibfield  {journal} {\bibinfo  {journal} {J. Fluid Mech.}\ }\textbf {\bibinfo {volume} {885}},\ \bibinfo {pages} {A45} (\bibinfo {year} {2020})}\BibitemShut {NoStop}%
\end{thebibliography}%

\end{document}